\documentclass[aps,prx,superscriptaddress,showpacs,reprint]{revtex4-1}
\usepackage{amsmath}
\usepackage{amssymb}
\usepackage{setspace}
\usepackage{graphicx}
\usepackage{subfigure}
\usepackage{braket}
\usepackage{mathrsfs}
\usepackage[colorlinks = true,linkcolor = red,urlcolor  = blue,citecolor = blue,anchorcolor = blue]{hyperref}

\begin{document}


\title{Phenomenology of soft gap, zero bias peak, and zero mode splitting in ideal Majorana nanowires}

\author{Chun-Xiao Liu, F. Setiawan, Jay D. Sau, S. Das Sarma}
\affiliation{Condensed Matter Theory Center and Joint Quantum Institute and Station Q Maryland, 
Department of Physics, University of Maryland, College Park, Maryland 20742-4111, USA}

\date{\today}

\begin{abstract}
We theoretically consider the observed soft gap in the proximity-induced superconducting state of semiconductor nanowires in the presence of spin-orbit coupling, Zeeman spin splitting, and tunneling leads, but in the absence of any extrinsic disorder (i.e., an ideal system).  We critically consider the effects of three distinct intrinsic physical mechanisms (tunnel barrier to normal leads, temperature, and dissipation) on the phenomenology of the gap softness in the differential conductance spectroscopy of the normal-superconductor junction as a function of spin splitting and chemical potential.  We find that all three mechanisms individually can produce a soft gap, leading to calculated conductance spectra qualitatively mimicking experimental results. We also show through extensive numerical simulations that the phenomenology of the soft gap is intrinsically tied to the broadening and the height of the Majorana zero mode-induced differential conductance peak above the topological quantum phase transition point with both the soft gap and the quality of the Majorana zero mode being simultaneously affected by tunnel barrier, temperature, and dissipation. We establish that the Majorana zero mode splitting oscillations can be suppressed by temperature or dissipation (in a similar manner), but not by the tunnel barrier.  Since all three mechanisms (plus disorder, not considered in the current work) are likely to be present in any realistic nanowires, discerning the effects of various mechanisms is difficult, necessitating detailed experimental data as a function of all the system parameters, some of which (e.g., dissipation, chemical potential, tunnel barrier) may not be known experimentally. While the tunneling-induced soft-gap behavior is benign with no direct adverse effect on the Majorana topological properties with the zero-bias peak remaining quantized at $2e^2/h$, the soft gap induced by finite temperature and/or finite dissipation is detrimental to topological properties and must be avoided as much as possible.
\end{abstract}

\maketitle


\section{introduction}
An ideal Majorana nanowire, as in the Kitaev spinless $p$-wave superconducting one-dimensional (1D) lattice chain model~\cite{Kitaev2001Unpaired}, is a topological superconductor wire with non-Abelian Majorana zero modes (MZMs) localized at the two wire ends. A continuum (and spinful) realistic version of this system was proposed in 2010~\cite{Lutchyn2010Majorana, Oreg2010Helical, Sau2010Non}, where the topological superconductivity is induced in a spin-orbit-coupled semiconductor wire with the superconducting pairing introduced by proximity effect using a neighboring regular metallic $s$-wave superconductor.  Theoretically, the nanowire enters the topological phase, with localized MZMs at the two wire ends, through a magnetic field tuned topological quantum phase transition (TQPT) when the Zeeman spin splitting ($V_Z$) in the nanowire exceeds ($V_Z>V_{Zc}$) a critical field defined by $V_{Zc}= \sqrt{\Delta^2 + \mu^2}$, where $\Delta$ is the induced superconducting gap in the nanowire (at zero field) and $\mu$ is the nanowire chemical potential.  The system acquires a topological gap with the gap size being proportional to the spin-orbit coupling strength $\alpha$ for $V_Z>V_{Zc}$ whereas for $V_Z<V_{Zc}$ the system is a non-topological (or trivial) superconductor.  Note that the superconducting gap in the nanowire vanishes precisely at the TQPT in the thermodynamic limit (infinite wire length) as is necessary for a TQPT separating a trivial and a topological phase~\cite{Read2000Paired}. It is known that such MZMs obey non-Abelian braiding statistics as long as the two MZMs are far enough from each other so that they can be considered isolated providing the system the so-called `exponential topological protection'~\cite{Read2000Paired, Kitaev2001Unpaired, Nayak2008Non-Abelian}. These non-Abelian MZMs can be used for topological quantum computation and have attracted great interest from physicists, mathematicians, engineers, and computer scientists over the last 15 years~\cite{Nayak2008Non-Abelian}.

The semiconductor nanowire proposal for artificially creating topological superconductivity and MZMs has attracted tremendous attention since the individual ingredients to experimentally realize this system in the laboratory are simple: proximity-induced superconductivity in a nanowire, spin-orbit coupling, and a magnetic field to induce Zeeman spin splitting~\cite{DasSarma2015Majorana, Alicea2012New, Beenakker2013Search, Leijnse2012Introduction, Stanescu2013Majorana, Elliott2015Colloquium}. Indeed a large number of experiments carried out by different groups during the last five years have led to experimental transport signatures in semiconductor nanowires claimed to be consistent with the existence of MZMs although none of these experiments provides definitive evidence yet~\cite{Mourik2012Signatures, Deng2012Anomalous, Das2012Zero, Churchill2013Superconductor, Finck2013Anomalous, Albrecht2016Exponential, Zhang2016Ballistic, Deng2016Majorana, Chen2016Experimental}.

It is essentially universally agreed that an ideal semiconductor-superconductor hybrid structure satisfying the pristine conditions proposed in the original theoretical papers~\cite{Lutchyn2010Majorana, Oreg2010Helical, Sau2010Generic, Sau2010Non, Lutchyn2011Search, Stanescu2011Majorana} should, by construction, possess MZMs, but it is not manifestly clear that the corresponding experimental systems are exact analog simulators of the ideal proposed Majorana nanowires. One obvious difference is, of course, the invariable presence of disorder, which is always present in the experimental systems. Since topological superconductivity is strongly adversely affected by disorder, complications arising from disorder in the nanowire have been extensively studied in this context~\cite{Brouwer2011Topological, Brouwer2011Probability, Kells2012Near, Sau2012Experimental, Takei2013Soft, Sau2013Density, Stanescu2011Majorana, Stanescu2013Dimensional, Hui2015Bulk, Cole2016Proximity, Adagideli2014Effects, Akhmerov2011Quantized}. It is well-accepted that the existence of fermionic subgap states, induced by disorder in the nanowire or the parent superconductor or the normal leads, leading to the appearance of a soft gap is disastrous to the topological properties of the Majorana modes. The current theoretical work leaves out all effects of disorder (and hence refers to an `ideal' system) since spectacular recent materials and fabrication developments have led to semiconductor-superconductor hybrid nanowire systems which have little disorder and are in the ballistic transport regime where the electrons travel essentially through the whole nanowire without any disorder scattering~\cite{Chang2015Hard, Deng2016Majorana, Gul2017Hard, Zhang2017unknown, Zhang2016Ballistic}. Therefore, it is now possible to discuss the properties of disorder-free ideal nanowires in the context of actual experimental systems.

The specific problem being studied in the current work is often referred to as the `soft gap' problem in the context of Majorana nanowire research activity.  Experimental differential conductance spectroscopy measures the differential conductance (i.e., $dI/dV$) as a function of a bias voltage $V$ applied across a normal metal (N)-superconducting nanowire (S) junction with an increasing applied magnetic field, finding a zero-bias conductance peak (ZBCP) forming above a finite magnetic field which is tentatively identified as the critical field $V_{Zc}$ for the theoretically predicted TQPT in the system. The signature for the MZM formation in the nanowire is taken as the development of the ZBCP above a critical magnetic field, as predicted theoretically a while ago~\cite{Sengupta2001Midgap, Law2009Majorana,Akhmerov2009Electrically, Flensberg2010Tunneling, Sau2010Non}. The soft gap issue relates to the fact that the nanowire superconducting gap as manifested in the differential conductance spectroscopy (i.e. $dI/dV$ measured as a function of $V$) of the NS junction appears to be soft with the conductance value being finite (and large, $\sim e^2/h$) within the gap instead of being hard (i.e. conductance almost vanishing inside the gap, $\ll e^2/h$). The gap actually appears to soften universally (i.e. in all the experimental studies) with increasing magnetic field, and the background conductance is often quite large by the time the ZBCP develops. In fact, in most experimental observations, the ZBCP rises from an essentially constant (and large) background conductance where the superconducting gap is not easy to identify. Concurrently, the ZBCP is always broad covering essentially half or more of the gap that may still exist for $V_Z>V_{Zc}$. Originally, this soft gap phenomenology was attributed to disorder, and it was even suggested, given the ubiquity and the severity of the gap softness, that perhaps the observed ZBCP has nothing whatsoever to do with MZMs, but is just a manifestation of disorder-induced weak anti-localization peak in the presence of spin-orbit coupling, the so-called class D disorder peaks~\cite{Sau2013Density, Bagrets2012Class, Liu2012Zero, Pikulin2012Zero}.  Recent experimental work involving epitaxial semiconductor-superconductor structures~\cite{Chang2015Hard, Deng2016Majorana, Gul2017Hard, Zhang2017unknown, Zhang2016Ballistic}, following the suggestion by Takei \textit{et al}.~\cite{Takei2013Soft}, has, however, eliminated disorder as the only possible source of the soft gap problem, since the soft gap itself still persists, at least for finite magnetic field values even in these essentially disorder-free ballistic nanowires. We define soft (hard) gap throughout this work purely empirically with the gap being soft (hard) depending entirely on the conductance ($G_S=dI/dV$) within the superconducting gap region being large (small) with $G_S \ll e^2/h$ throughout the gap being our definition of a hard gap. If this low subgap conductance condition is violated, the induced gap is considered `soft'.

In order to better understand the persistence of a soft induced gap in Majorana nanowires even in the absence of any obvious extrinsic disorder~\cite{Deng2016Majorana, Gul2017Hard, Zhang2017unknown}, particularly at higher magnetic field, we undertake in the current work a detailed investigation of the conductance spectra in nanowire NS junctions using a minimal model, which includes only proximity-induced $s$-wave superconductivity $\Delta$, spin-orbit coupling, and tunable Zeeman splitting induced by an external magnetic field. We keep only one spinful conduction channel (``subband") in the nanowire without any loss of generality since our interest is only in understanding the qualitative dependence of the soft induced gap and the ZBCP on various system parameters.  Our goal is to develop understanding of how a soft gap can develop in proximity-induced nanowires in the absence of any disorder and how it evolves with increasing magnetic field, eventually going through the TQPT and forming the ZBCP.  We want to understand the interplay of the soft gap with the ZBCP formation as a function of various intrinsic system parameters. The NS junction itself is modeled by an effective delta-function tunnel barrier with a strength of $Z$ to keep things tractable. The other intrinsic parameters of the system are the nanowire chemical potential $\mu$, the wire length $L$ and temperature $T$. In addition to these seven obvious experimentally relevant system parameters ($\Delta$, $\alpha$, $V_Z$, $Z$, $\mu$, $L$, $T$) we introduce one other parameter following Refs.~\cite{DasSarma2016How, Liu2017Role}, $\Gamma$, which is a dissipative broadening in the system arising from some unknown origin. This dissipation term is often necessary to bring experiment and theory into qualitative agreement~\cite{DasSarma2016How, Liu2017Role}, and we use $\Gamma$ as a phenomenological parameter in the theory. The origin of $\Gamma$ in the experimental Majorana nanowires is outside the scope of the current work, but we speculate that it could possibly arise from coupling to the parent superconductor providing the proximity effect and/or from vortices or quasiparticles invariably present in the environment. Essentially, $\Gamma$ simulates the coupling of the superconducting nanowire to a fermionic bath, which is often unavoidable in experimental situations~\cite{DasSarma2016How}. Although $\Gamma$ may be intrinsic or extrinsic depending on its physical origin, we keep $\Gamma$ in our theory as an adjustable parameter to show its effect on the soft gap phenomena. In some sense, $\Gamma$ should be thought of as incorporating in a single phenomenological parameter all the extrinsic effects left out of our model. Obviously, $\Gamma$ must be small ($\Gamma \ll \Delta$) for the nanowire to manifest topological behavior~\cite{DasSarma2016How}.

We note that even for fixed induced gap ($\Delta$), spin-orbit coupling ($\alpha$), wire length ($L$), and number of channels in the nanowire (just one spinful subband in the model) there are 5 independent parameters (all with dimensions of energy) in our theory ($V_Z, Z, \mu, T, \Gamma$), making the problem rather challenging since our goal is to figure out how the phenomenology of soft gap as well as of ZBCP evolves as a function of all five variables. A five-dimensional functional behavior is difficult to visualize, but the experimental results are likely to depend on all five of these variables, necessitating our presenting a great number of results to provide a quantitative and qualitative feel for the soft gap phenomenology as a function of these five important experimental variables.  In principle, all five of these parameters should be known in the experiment, but in practice none of them may actually be known quantitatively, making direct comparison between experiment and theory essentially impossible. For example, although the applied magnetic field is obviously known in the experiment, the corresponding Zeeman splitting energy is not known since the Lande $g$-factor is likely to be strongly renormalized (and field-dependent)~\cite{Cole2015Effects} in the actual experimental situation and has not been measured in the Majorana nanowire system. Similarly, the NS tunnel barrier strength $Z$ can to some extent be controlled by a suitable gate which could raise (lower) the barrier, thus decreasing (increasing) tunneling amplitude, but its absolute magnitude is unknown. The chemical potential in the wire could in principle be measured in isolated nanowires, but it is unknown in the experimental situation with the nanowire lying on a superconductor although the chemical potential could possibly be varied somewhat by a suitable gate potential applied directly on the wire. The base temperature ($\sim 50~$mK typically in a dilution refrigerator) of the system is known, but the electron temperature is likely to be considerably higher and is unknown. Finally, the dissipation parameter $\Gamma$ is unknown by definition which can only be estimated by its effect on the measured tunneling spectra. What is even worse from the standpoint of quantitative understanding of the experimental data is that even the fixed parameters (induced gap, spin-orbit coupling, wire length, effective number of occupied subbands) of the theory are unknown experimentally: the induced gap is known approximately, but not precisely, from the experimental tunneling spectra; the spin-orbit coupling is known at best in isolated wires whereas a very substantial modification is expected in the experimental condition lying in contact with a parent superconductor; the number of occupied subbands in contact with the superconductor cannot really be measured directly at all;  and finally, although the nominal wire length is known, the effective length is unknown since we do not know the precise extent over which the superconductivity is proximity-induced.  Given all these complications about unknown experimental parameters (and the large number of relevant parameters determining the tunnel conductance even within the minimal model), the best one can hope for is to develop a detailed qualitative physical understanding of the soft gap phenomenology on various system parameters.  This is what we aspire to in the current work.

An important question in the context of the observed soft gap (defined operationally in this work as $G_S \ll e^2/h$ hard gap criterion not being obeyed) in Majorana nanowires is whether it is detrimental to the MZM non-Abelian properties, i.e., whether soft gap directly suppresses the topological behavior. If so, then the soft gap must always be eliminated for the ZBCP to have anything to do with the characteristic Majorana behavior.  On the other hand, if the soft gap is benign and does not adversely affect the MZM topological properties, there is no particular concern as long as MZMs can be clearly identified in the system. It is well-established that disorder-induced soft gap is detrimental to topological behavior since disorder produces many random MZMs along the wire leading to considerable MZM overlap destroying their topological immunity. But it is now experimentally clear that eliminating disorder does not always  necessarily lead to a hard gap~\cite{Zhang2016Ballistic, Deng2016Majorana, Gul2017Hard, Zhang2017unknown}, and therefore, we want to investigate the reasons for the soft gap in ballistic samples which are relatively disorder free. We find that all five of the experimental variables mentioned above ($V_Z, Z, \mu, T, \Gamma$) can lead to soft-gap behavior in different ways, and we provide extensive numerical results showing the evolution of the soft gap with system parameters. The intrinsic soft gap arising from the low tunnel barrier effect is, however, benign since it does not adversely affect any topological properties. Much of the current work is aimed at understanding how the tunneling itself could lead to the soft-gap behavior in the nanowire as observed experimentally with the system manifesting a true hard gap only in the extreme weak-tunneling limit where the tunnel barrier $Z$ is very large. Such a tunneling-induced soft gap is not detrimental to topological properties.

Equally important with the soft-gap problem, we also study concurrently the phenomenology of the ZBCP quality including its height and broadening as well as its splitting behavior above the TQPT where the overlap between the two localized MZMs at the two ends should produce MZM oscillations in the system. The question arises now how various parameters producing the soft gap affect the ZBCP properties associated with the MZM. This is an important question because experimentally it is often found that ZBCP properties correlate with the soft-gap behavior in specific ways. We emphasize in this context that by construction our ideal Majorana nanowire model can only have ZBCPs arising from MZMs since we do not include extrinsic disorder or Andreev bound state effects~\cite{Liu2017Andreev, Stenger2017Tunneling}.  Thus, our results for soft gap also directly give us the parameter-dependence of the MZM-induced ZBCP as $V_Z$ is increased above $V_{Zc}$.

We have used two complementary techniques (within the same minimal model) to obtain our results in order to make the underlying physics crystal clear. The first method (Sec.~\ref{sec:semiinfinite}) uses the original Blonder-Tinkham-Klapwijk (BTK) theory~\cite{Blonder1982Transition} for a semiinfinite NS junction without any dissipation term to obtain the parameter dependence of soft gap and ZBCP on $V_Z, Z, \mu$ and $T$. This is done to definitively show that decreasing the tunnel barrier strength $Z$ immediately leads to a soft-gap behavior which in fact becomes stronger with increasing $V_Z$ (i.e., the gap softens further with increasing $V_Z$ for a fixed $Z$). These findings are in agreement with experiment, and establish that the soft gap may arise entirely from the system being away from the high-tunnel-barrier extreme tunneling limit. This tunneling-induced soft-gap behavior (i.e., a soft gap developing purely by virtue of the system being away from the weak-tunneling limit) may be referred to as the `BTK-soft-gap' since it is arising from BTK-type physics at the nanowire-normal lead tunnel junction. In order to quantify the barrier-induced gap softening, we compare the subgap conductance $G_S$ with the corresponding above-gap normal conductance $G_N$, showing that the expected $G_S \sim G_N^2$ dependence holds in the extreme tunneling limit ($G_N \ll e^2/h$) at $T=0$.  We also test the extent to which the Andreev conductance formula derived by Beenakker~\cite{Beenakker1992Transport} for single-subband transport holds in this case, finding that the Beenakker formula holds for $V_Z=0$ and large $\mu$ (which are the conditions under which the Beenakker single-subband formula was derived).  At finite field, however, we find that the gap always softens faster than that indicated by the Beenakker formula in agreement with a recent experimental finding~\cite{Zhang2017unknown}. Finite temperature effects soften the gap further due to thermal excitation of particle-hole pairs introducing considerable broadening in the MZM-induced ZBCP. The second technique we employ (Sec.~\ref{sec:realistic}) uses a more realistic numerical simulation for finite length wires with a dissipative broadening $\Gamma$ being present in the theory. The dependence of the gap softening and ZBCP broadening due to tunnel barrier and temperature variations is similar here as in the first technique with the gap softening strongly with increasing $V_Z$ and decreasing $Z$ as before. In addition, temperature both softens the gap further and broadens the ZBCP. The effect of dissipation $\Gamma$ on the gap softening and ZBCP broadening are qualitatively similar. Above the TQPT ($V_Z>V_{Zc}$), both temperature and dissipation tend to suppress the MZM oscillations in the ZBCP, but decreasing $Z$ does not suppress MZM oscillations although it softens the gap and broadens the ZBCP. Thus, MZM oscillations should  be present in Majorana nanowires even if $Z$ is relatively small (and consequently the gap very soft) as long as both temperature and dissipation are small in the system. Thus, BTK physics itself, even in the strong-tunneling limit, cannot suppress the MZM oscillations expected in the ZBCP. Given that MZM oscillations are rarely experimentally observed in Majorana nanowires, we conclude that effective temperature and/or effective dissipation are not yet very small in the experimental systems, even if the soft gap itself may arise (at least partially) from BTK physics (i.e., strong tunneling).

The rest of this paper is organized as follows. In Sec.~\ref{sec:model} we describe our model, explain the minimal theory and how the numerical calculations are carried out for obtaining differential conductance $G=dI/dV$ for the NS junction involving Majorana nanowires. We also provide our parameter details in Sec.~\ref{sec:model}. In Sec.~\ref{sec:semiinfinite} we provide and discuss our results for the semiinfinite BTK model calculation of differential conductance (without any dissipation) as a function of $Z, V_Z, \mu$ and $T$. We provide detailed results of subgap conductance $G_S$ as a function of normal conductance $G_N$ in this section comparing with various theoretical predictions. In Sec.~\ref{sec:realistic}, we consider realistic finite length nanowires with dissipation and provide our results for differential conductance as a function of $Z, V_Z, \mu, T, L$ and $\Gamma$. We discuss effects of these parameters on gap softening, ZBCP height and width, and MZM oscillations pointing out qualitative agreement (or not) with existing experimental results.  We conclude in Sec.~\ref{sec:discussion} discussing the implications of our results in the currently ongoing search for non-Abelian MZMs in semiconductor nanowires and considering several open questions.


\section{Model}\label{sec:model} 

We begin by considering a 1D NS junction between a spin-orbit-coupled normal lead with a spin-orbit-coupled nanowire (NW) in proximity to an $s$-wave superconductor~\cite{Lutchyn2010Majorana, Oreg2010Helical}. Following the BTK paper~\cite{Blonder1982Transition}, we model the barrier at the junction as a delta-function barrier with a strength $Z$. In the particle-hole space, the Hamiltonian can be written as 
\begin{align}
H_j = \int dx\Psi^{\dagger}_j(x)\mathcal{H}_j(x)\Psi_j(x),
\end{align}
where $\Psi_j(x) = (\psi_{j\uparrow}(x),\psi_{j\downarrow}(x),\psi^\dagger_{j\downarrow}(x),-\psi^\dagger_{j\downarrow}(x))^{\mathrm{T}}$ are the Nambu spinors with $\psi^\dagger_{j\sigma}(x)$ ($\psi_{j\sigma}(x)$) being the creation
(annihilation) operator for an electron with spin $\sigma$ in
region $j= \mathrm{lead}$  and $\mathrm{NW}$. The Bogouliubov-de Gennes (BdG) Hamiltonian of the lead and nanowire is given by
\begin{subequations}\label{eq:hamiltonian}
\begin{align}
\mathcal{H}_{\mathrm{lead}} &=  \left(-\frac{\hbar^2\partial_x^2}{2m} - \mu_{\mathrm{lead}}\right)\tau_z -i \alpha \partial_x \tau_z\sigma_y + V_Z \sigma_x , \\
\mathcal{H}_{\mathrm{NW}} &= \left(-\frac{\hbar^2\partial_x^2}{2m} - \mu\right)\tau_z -i \alpha \partial_x \tau_z\sigma_y + V_Z \sigma_x + \Delta \tau_x,\label{eq:H_NW}
\end{align}
\end{subequations}
where $\mu_{\mathrm{lead}}$ ($\mu$) is the chemical potential of the lead (nanowire),
$\alpha$ is the spin-orbit coupling strength, $V_Z$ is the Zeeman field, $\Delta$ is the proximity-induced $s$-wave pairing potential, and
$\tau_{x,y,z}$ ($\sigma_{x,y,z}$) are the Pauli matrices acting on the particle-hole
(spin) subspace. For simplicity, in the following we will set $\hbar = 1, \Delta = 1$, the Boltzman constant $k_B = 1$ and the spin-orbit length $l_{SO}= \hbar^2/(m\alpha) = 1$, where $m=0.015m_e$ and $\alpha$ = 0.5 eV{\AA}.  We use  $\mu_{\mathrm{lead}}= 25$ for the numerical simulation in this paper. Varying $\mu_{\mathrm{lead}}$ has no qualitative effect on any of the presented results. For finite nanowires, certain amount of dissipation can be added to the nanowire Hamiltonian, i.e., Eq.\eqref{eq:H_NW}, in the form of $i\Gamma$~\cite{DasSarma2016How, Liu2017Role}. This dissipation, which is typically small, most likely arises from vortices in the parent superconductor (and perhaps also in the nanowire itself). The details of dissipation in this context have already been discussed in depth in Refs.~\cite{DasSarma2016How, Liu2017Role}.

In this paper, we calculate the differential conductance $G = dI/dV$ for the case where the nanowire is semiinfinite (Sec.~\ref{sec:semiinfinite}) and for a more realistic case where the nanowire has a finite length (Sec.~\ref{sec:realistic}). We numerically calculate the conductance by discretizing the Hamiltonian (Eq.~\eqref{eq:hamiltonian}) into a 1D lattice and obtaining the scattering matrix~\cite{Setiawan2015Conductance} from the numerical transport package Kwant~\cite{Groth2014Kwant}. The zero-temperature conductance (given in the unit of $e^2/h$) is computed using the following formula 
\begin{equation}\label{eq:conductance}
G_0 = 2 + \sum_{\sigma,\sigma' = \uparrow \downarrow} \left(|r^{\sigma \sigma'}_{eh}|^2 -  |r^{\sigma \sigma'}_{ee}|^2 \right), 
\end{equation}
where $r_{eh}$ and $r_{ee}$ are the Andreev and normal reflection amplitudes, respectively. The factor of 2 in Eq.~\eqref{eq:conductance} arises from the fact that we consider a one-subband system with two spin channels. The finite-temperature conductance is calculated from the zero-temperature conductance by a convolution with the derivative of the Fermi function $f$, i.e.,
\begin{equation}\label{eq:GT}
G_T(V) = - \int_{-\infty}^{\infty} dE G_0(E)\frac{df(E-V)}{dE}.
\end{equation}
Since the theoretical methods used in the current work are standard and have been discussed extensively in the literature~\cite{Sau2010Non, Stanescu2011Majorana, DasSarma2016How, Liu2017Role, Liu2017Andreev,Setiawan2015Conductance,Lin2012Zero}, we do not provide any further details on the theory, focusing instead on the numerically calculated conductance results.

\section{Semiinfinite Nanowire}\label{sec:semiinfinite}
In this section, we give the result for the case where the nanowire is semiinfinite. In the following subsections, we will systematically compare the differential conductance $dI/dV$ for different tunnel barrier strength $Z$, Zeeman field $V_Z$ and temperature $T$. In particular, we will focus on the effects of these three parameters on the subgap conductance.

\subsection{Zero temperature}

\begin{figure}[h]
\begin{center}
\includegraphics[width=\linewidth]{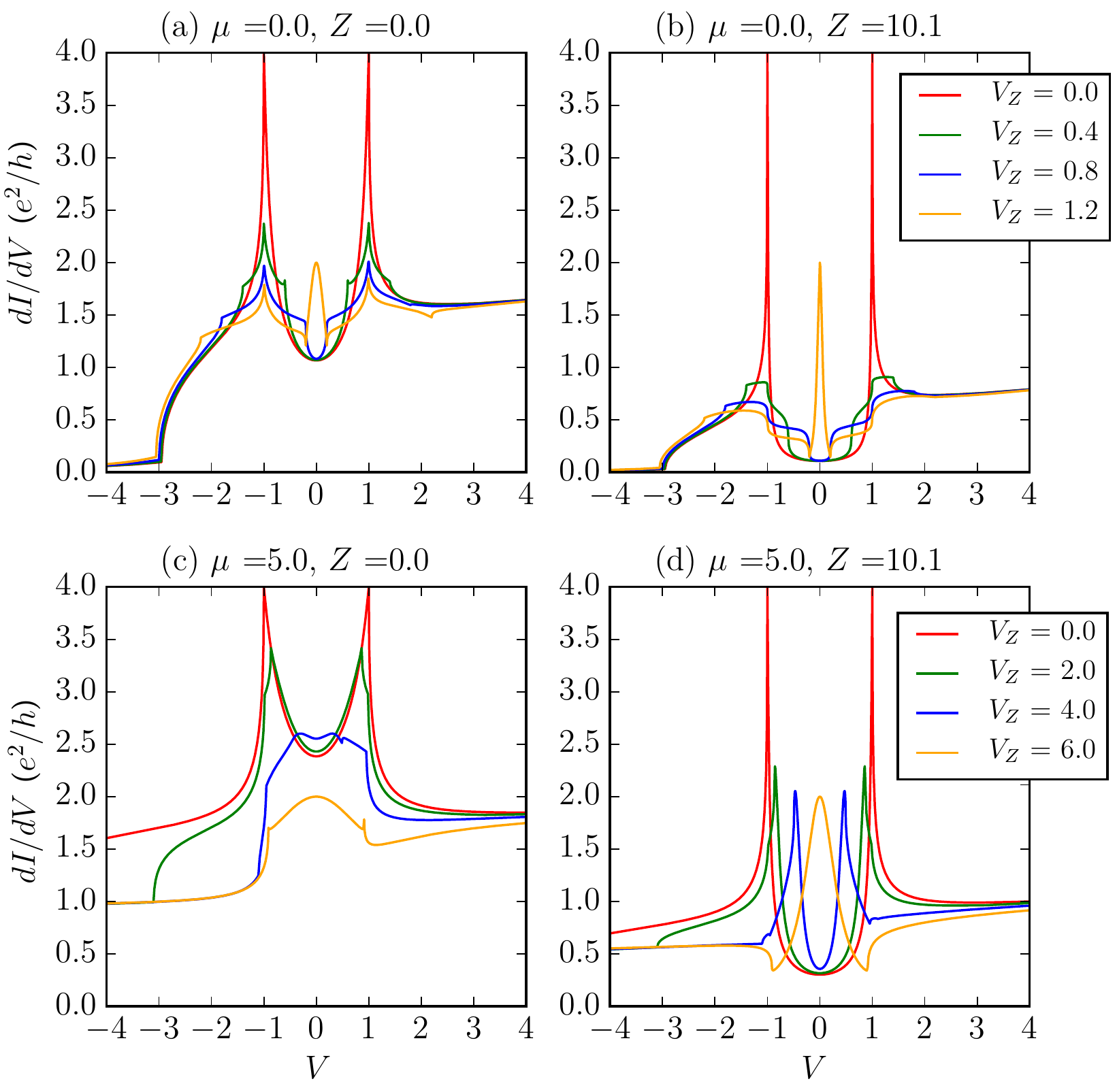}
\end{center}
\caption{(color online) Zero-temperature differential conductance $dI/dV$ vs voltage $V$ for different chemical potential [$\mu = 0$ (top panel) and $\mu = 5$ (bottom panel)] and barrier strength [$Z = 0$ (left panel) and $Z = 10.1$ (right panel)]. The Zeeman field strengths in all the panels are below the TQPT except the largest Zeeman field.  The value of $\Delta$ is 1.}\label{fig:dIdV_barrier} 
\end{figure}

Figure~\ref{fig:dIdV_barrier} shows the calculated zero-temperature differential conductance of the junction for different Zeeman field $V_Z$, chemical potential $\mu$ and tunnel barrier strength $Z$. The in-gap conductance is always particle-hole symmetric due to the unitarity of the in-gap reflection matrix. However, the above-gap conductance can be particle-hole asymmetric. As shown in Fig.~\ref{fig:dIdV_barrier}, this particle-hole asymmetry is more pronounced for small barrier strength or small chemical potential. 

By comparing the top and bottom panels in Fig.~\ref{fig:dIdV_barrier}, we can also see that the conductance generally decreases with increasing barrier strength $Z$ except the zero-bias conductance value in the topological region which is always quantized at $2 e^2/h$~\cite{Law2009Majorana, Flensberg2010Tunneling, Wimmer2011Quantum, Setiawan2015Conductance} for $V_Z > V_{Zc}$ independent of the background subgap conductance. The coherence peak at zero temperature and zero Zeeman field is always quantized at $4 e^2/h$ due to equal weight of the particle and hole-component of the BCS wave function at the gap edge (i.e., $\sum_{\sigma = \uparrow,\downarrow}|{u}_{\sigma}|^2 = \sum_{\sigma = \uparrow,\downarrow}|{v}_{\sigma}|^2$). Below the TQPT, as $V_Z$ is raised, the gap shrinks with the coherence peak strength decreasing and the subgap conductance increasing. This increase in the subgap conductance below the TQPT as $V_Z$ is raised can be seen more clearly in the $dI/dV$ plot for larger $\mu$ (see bottom panel of Fig.~\ref{fig:dIdV_barrier}) because for larger $\mu$, there is a larger range of $V_Z$ below the critical value, $V_{Zc} = \sqrt{ \Delta^2 + \mu^2 }$~\cite{Lutchyn2010Majorana, Oreg2010Helical, Sau2010Non}, at which the TQPT happens. As $V_Z$ is raised past the TQPT, the ZBCP (quantized at $2e^2/h$) appears which indicates the presence of an  MZM at the end of the nanowire. The width of the ZBCP increases with decreasing barrier strength $Z$~\cite{Lin2012Zero, Setiawan2015Conductance}. For the case where the tunnel barrier strength is not sufficiently high, the ZBCP width can span the entire gap (with a substantial amount of subgap conductance) and the superconducting gap may not appear as peaks in the conductance profile (see Fig.~\ref{fig:dIdV_barrier}). This `soft gap' phenomenon, which is associated with increasing Zeeman field, occurs generically in the experiment~\cite{Deng2016Majorana, Gul2017Hard, Zhang2017unknown} which probably indicates that the tunnel barrier strength used in the experiment is not sufficiently high to suppress the subgap conductance. We would like to point out that unlike the disorder-induced soft gap, the BTK soft-gap behavior that arises due to small tunnel barrier strength is not detrimental to the non-Abelian property of the MZM~\cite{DasSarma2016How}.

\begin{figure}[h]
\begin{center}
\includegraphics[width=\linewidth]{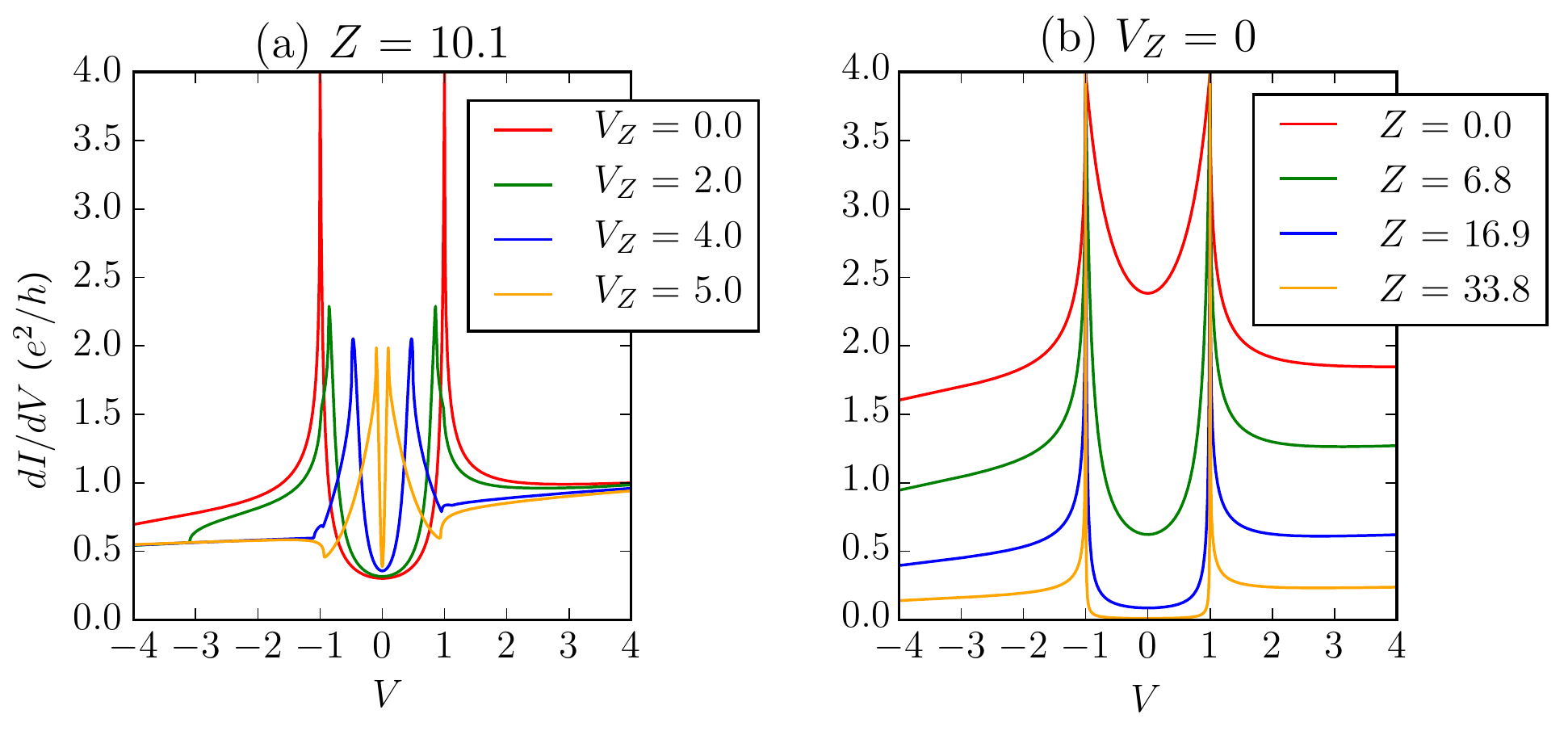}
\end{center}
\caption{(color online) Zero-temperature differential conductance $dI/dV$ vs voltage $V$ for chemical potential $\mu = 5$. (a) $dI/dV$ for barrier strength $Z = 10.1$ with different Zeeman field strength below the TQPT. (b) $dI/dV$ for $V_Z = 0$ and different barrier strength. The value of $\Delta$ is taken to be 1.}\label{fig:dIdV_VZ_barrier} 
\end{figure}

Figure~\ref{fig:dIdV_VZ_barrier} shows more explicitly the dependence of the conductance on the barrier strength and the Zeeman field strength below the TQPT. Again we can see from Fig.~\ref{fig:dIdV_VZ_barrier}(a) that as the Zeeman field increases towards the TQPT, the gap shrinks and the subgap conductance increases. Fig.~\ref{fig:dIdV_VZ_barrier}(b) shows the dependence of conductance on the tunnel barrier strength. As can be seen in the figure, the conductance decreases with increasing barrier strength $Z$.

\begin{figure}[h]
\begin{center}
\includegraphics[width=\linewidth]{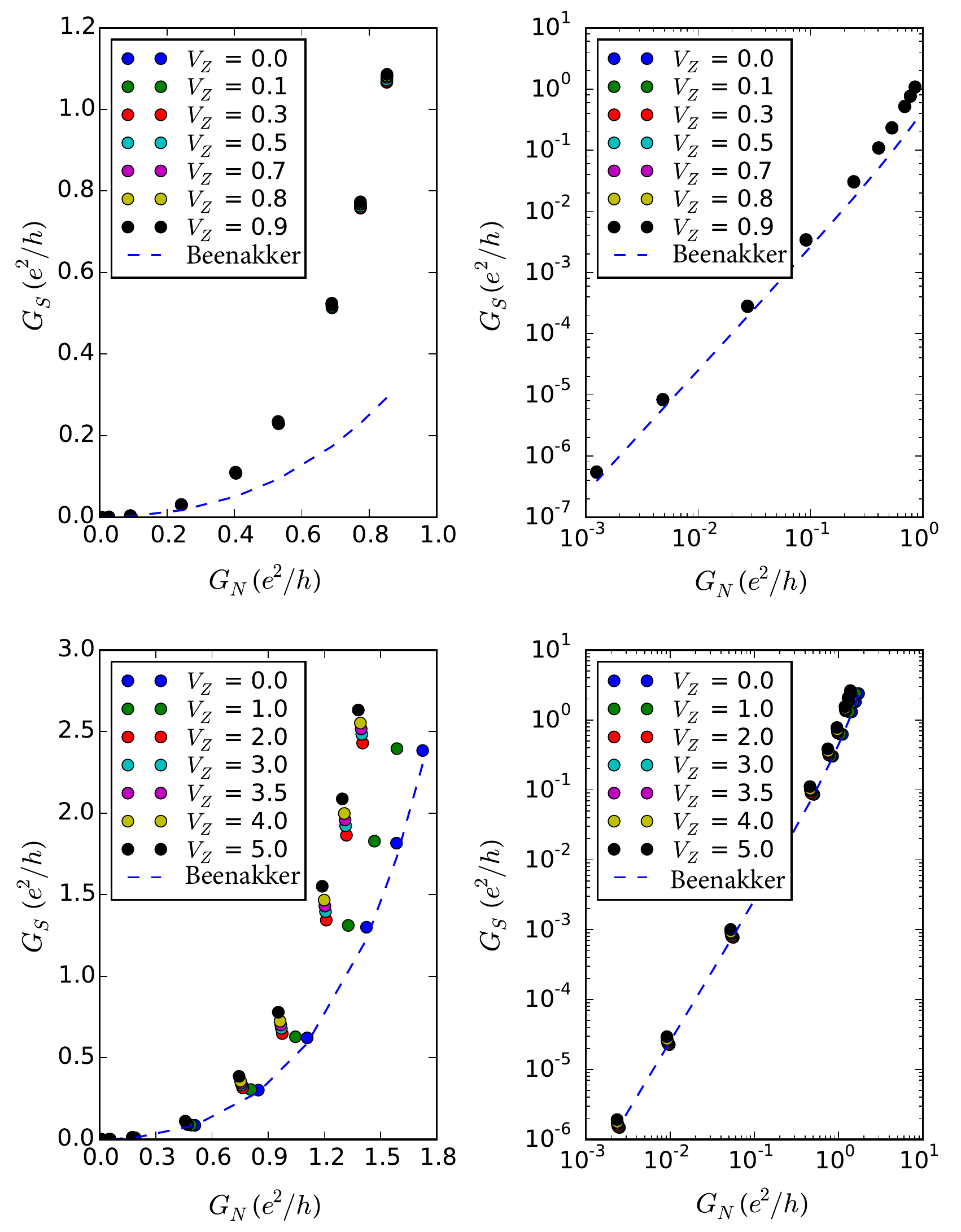}
\end{center}
\caption{(color online) Linear (left panel) and log-log plots (right panel) of zero-temperature values of $G_S$ vs $G_N$ for different Zeeman field $V_Z$ below the TQPT and different chemical potential [$\mu = 0$ (top panel) and $\mu = 5$ (bottom panel)]. The dashed line is the fit to the Beenakker formula $G_{S} = (G_N)^2/(2 - G_N/2)^2$. The Beenakker formula holds only for $V_Z = 0$ and $\mu \gg \Delta$. The value of $\Delta$ is taken to be 1.}\label{fig:GNGS} 
\end{figure}

\begin{figure}[h]
\begin{center}
\includegraphics[width=\linewidth]{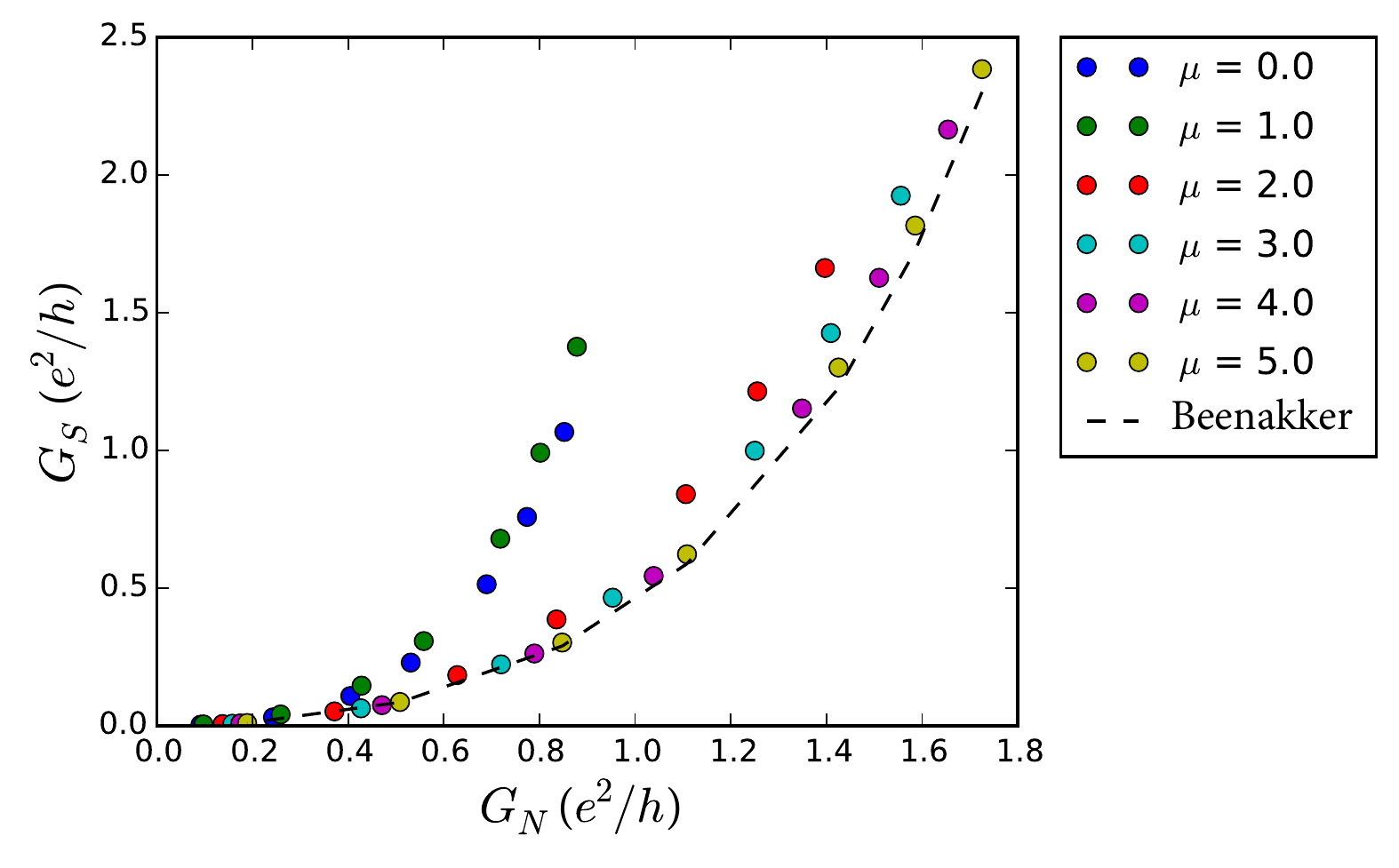}
\end{center}
\caption{(color online) Plot of zero-temperature values of $G_S$ vs $G_N$ for $V_Z = 0$ and different value of $\mu$.  The dashed line is the fit to the Beenakker formula $G_{S} = (G_N)^2/(2 - G_N/2)^2$. The Beenakker formula holds only for $\mu \gg \Delta$. The value of $\Delta$ is taken to be 1.}\label{fig:GNGS_diff_mu} 
\end{figure}

To quantify the hardness of the gap below the TQPT, we calculate the subgap conductance at zero voltage $G _S$. Throughout this paper, we define hard gap as $G_S \ll e^2/h$. The value of $G_S$ decreases with decreasing transparency (i.e., increasing $Z$). Since the junction transparency is determined not only by the tunnel barrier strength $Z$, but also by sharp variation of model parameters across the junction, i.e., the mismatch in the Fermi energy and spin-orbit coupling, we characterize the junction transparency by $G_N$, where $G_N$ is the normal-state conductance which is the above-gap conductance at large voltages. Since the above-gap conductance can in general be particle-hole asymmetric, the value of $G_N$ is taken to be 
\begin{equation}
G_N = \frac{G_{N+}+G_{N-}}{2},
\end{equation} 
where $G_{N\pm}$ is the conductance at large positive and negative voltages, respectively (for this paper, it is taken to be the conductance at $V = \pm 4 \Delta$, although other similar definitions do not have any qualitative effects). Our definitions for $G_S$ and $G_N$ are similar to those used in describing the experimental data~\cite{Zhang2016Ballistic, Gul2017Hard, Zhang2017unknown}.

\begin{figure}[h]
\begin{center}
\includegraphics[width=\linewidth]{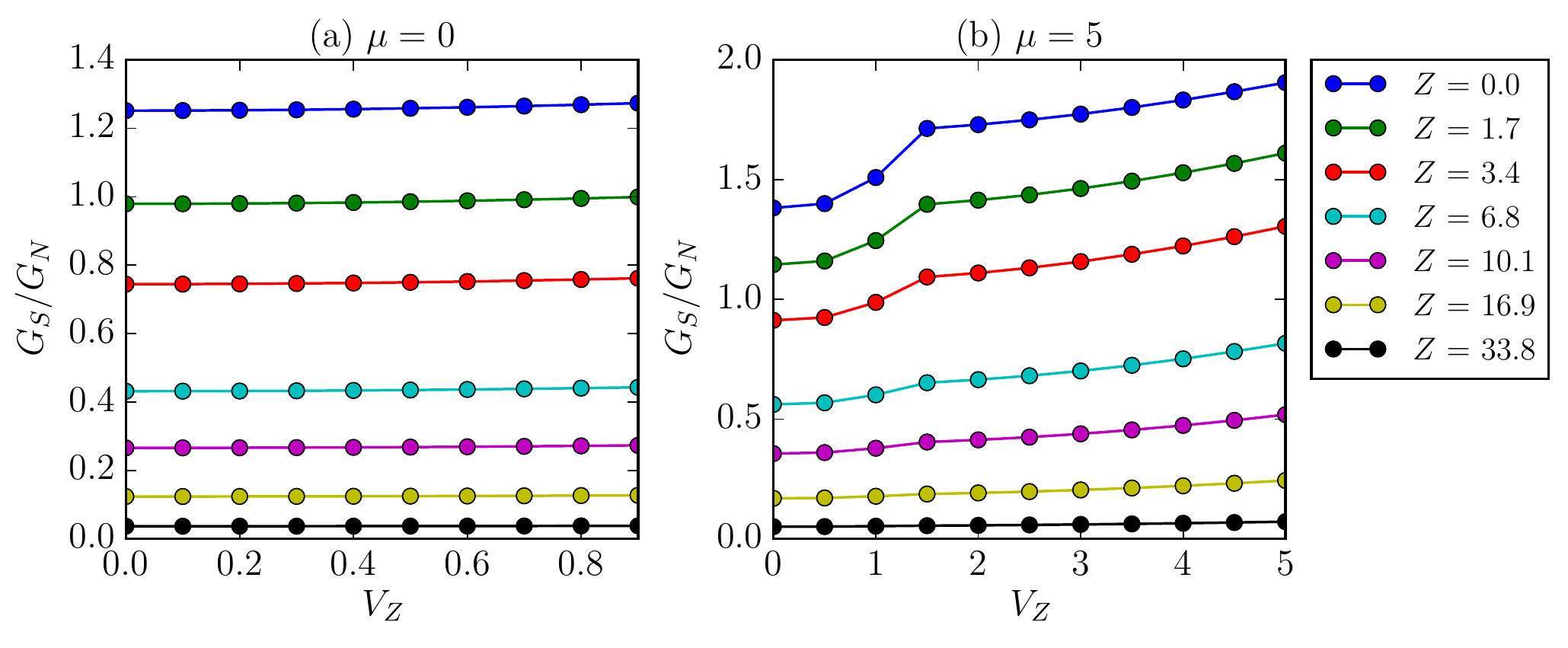}
\end{center}
\caption{(color online) Plot of zero-temperature values of $G_S/G_N$ vs $V_Z$ for different barrier strength $Z$ and different chemical potential [(a) $\mu = 0$ (top panel) and (b) $\mu = 5$ (bottom panel)]. The value of $V_Z$ is taken to be below the TQPT and the value of $\Delta$ is 1.}\label{fig:GNGS_diff_bh} 
\end{figure}

Figure~\ref{fig:GNGS} shows the linear and log-log plots of the zero-temperature $G_S$ vs $G_N$ values for several values of Zeeman field strength below the TQPT. For a given $V_Z$, each data point in Fig.~\ref{fig:GNGS} is obtained by varying the barrier height in the numerical simulation. For the same barrier height, the value of $G_S$ increases with increasing $V_Z$ which implies that the subgap conductance increases with increasing $V_Z$.  For large $\mu$ ($\mu \gg \Delta$, where Andreev appromixation holds) and zero $V_Z$, the value of $G_S$ is related to $G_N$ through the Beenakker formula $G_{S} = (G_N)^2/(2-G_N/2)^2$~\cite{Beenakker1992Transport} (see Figs.~\ref{fig:GNGS} and~\ref{fig:GNGS_diff_mu}), where $G_N/2$ is the transparency of each spin channel. This particular formula is derived for the single-subband situation, and has been used to discuss experimental data~\cite{Zhang2016Ballistic, Gul2017Hard, Zhang2017unknown}. As shown in Fig.~\ref{fig:GNGS}, for finite $V_Z$ the $G_S$ vs $G_N$ curve deviates from the Beenakker formula where for a given $G_N$, the subgap conductance $G_S$ is always greater than its counterpart at zero $V_Z$. This indicates that the gap softens as the Zeeman field increases, with the Beenakker formula defining the allowed gap hardness for a given transparency. This Zeeman-field-induced gap softening with the $G_S$ vs $G_N$ curve deviating from the Beenakker formula is in agreement with recent experimental observation~\cite{Zhang2017unknown}. Moreover, Fig.~\ref{fig:GNGS_diff_mu} shows that the $G_S$ vs $G_N$ curve can also deviate from the Beenakker formula even at zero $V_Z$ provided that the chemical potential is not too large. Figure~\ref{fig:GNGS_diff_bh} shows the ratio of $G_S/G_N$ vs $V_Z$ for different chemical potential $\mu$ and barrier strength $Z$. As shown in the figure, the effect of Zeeman field on the subgap conductance below the TQPT is more pronounced for larger chemical potential and smaller barrier strength. The most important qualitative feature of tunneling-induced soft gap is that the gap must harden in the weak-tunneling limit ($G_N \ll e^2/h$), where $G_S \sim G_N^2$ must hold. This is of course inherent in the Beenakker formula too, since the Beenakker formula is based on the BTK model. It should be emphasized that the ZBCP above the TQPT always has the quantized Majorana value $2e^2/h$ even when the gap is soft as long as the softening arises purely from tunneling effects. The topological properties of the nanowire are thus fully preserved even if the gap is very soft provided the softness arises from BTK tunneling effect.

\subsection{Effect of temperature}
\begin{figure}[h]
\begin{center}
\includegraphics[width=\linewidth]{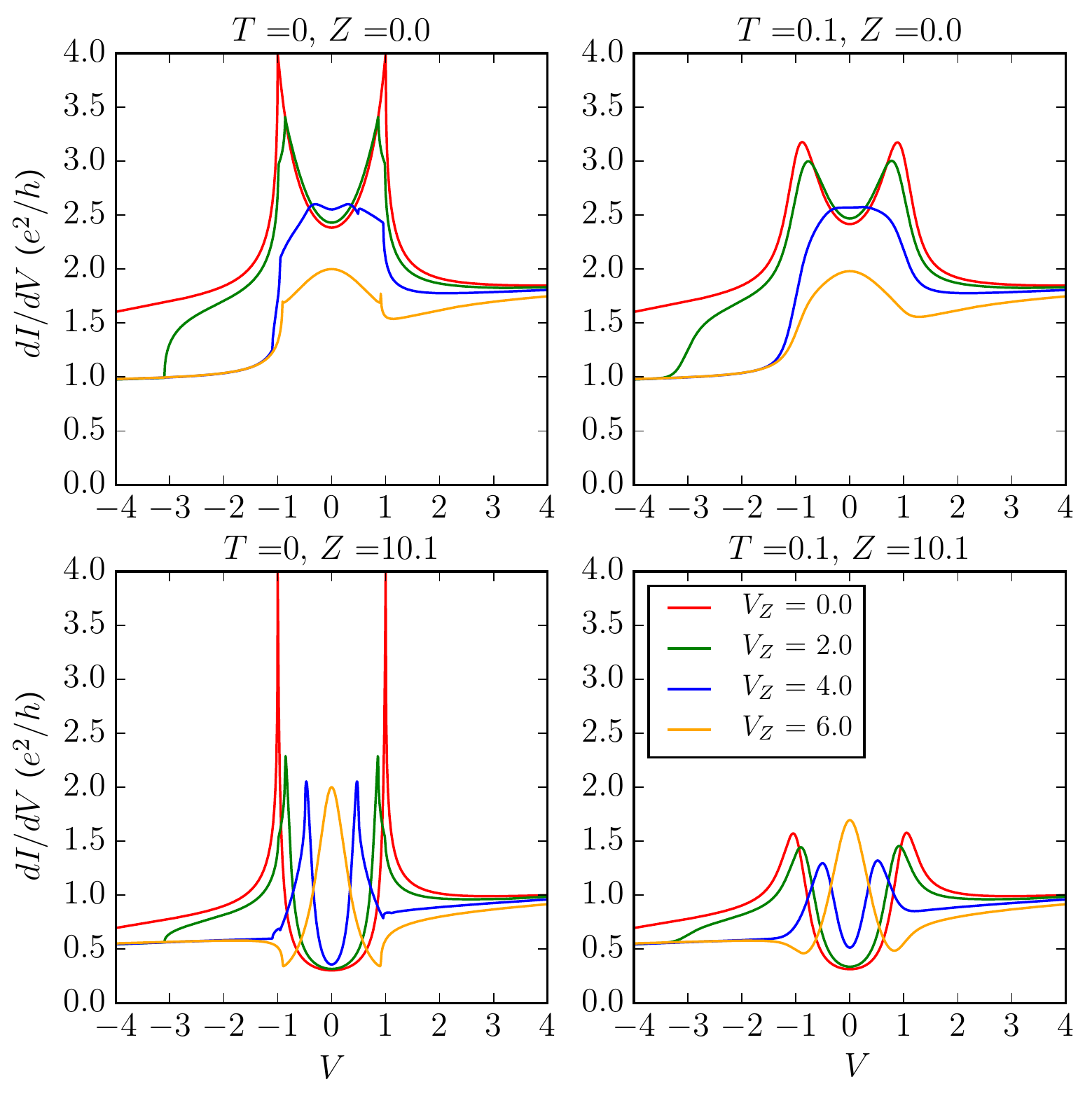}
\end{center}
\caption{(color online) Differential conductance $dI/dV$ vs $V$ for chemical potential $\mu = 5$ and different Zeeman fields $V_Z$, temperatures [$T = 0$ (left panel) and $T = 0.1$ (right panel)], and barrier strengths [$Z =0$ (top panel) and $Z = 10.1$ (bottom panel)]. All the Zeeman field strengths shown are below the QPT except the largest Zeeman field.  The value of $\Delta$ is taken to be 1.}\label{fig:dIdV_diff_temp} 
\end{figure}

Having discussed the effect of Zeeman field and barrier strength on the conductance at $T=0$, let us now look at the effect that the temperature has on the conductance. Figure~\ref{fig:dIdV_diff_temp} gives a comparison between the zero-temperature (left panel) and finite-temperature differential conductance (right panel) for different barrier strength and Zeeman field. Finite temperature broadens the coherence and zero-bias peaks and also lowers their peak values. At the same time, it increases the subgap conductance which gives rise to additional soft-gap behavior in the conductance spectrum, even if the tunneling gap itself is hard. Thus, finite temperature itself leads to gap softening, which could be considerable when the temperature in comparable to the gap.

\onecolumngrid

\begin{figure*}[h]
\begin{center}
\includegraphics[scale = 0.65]{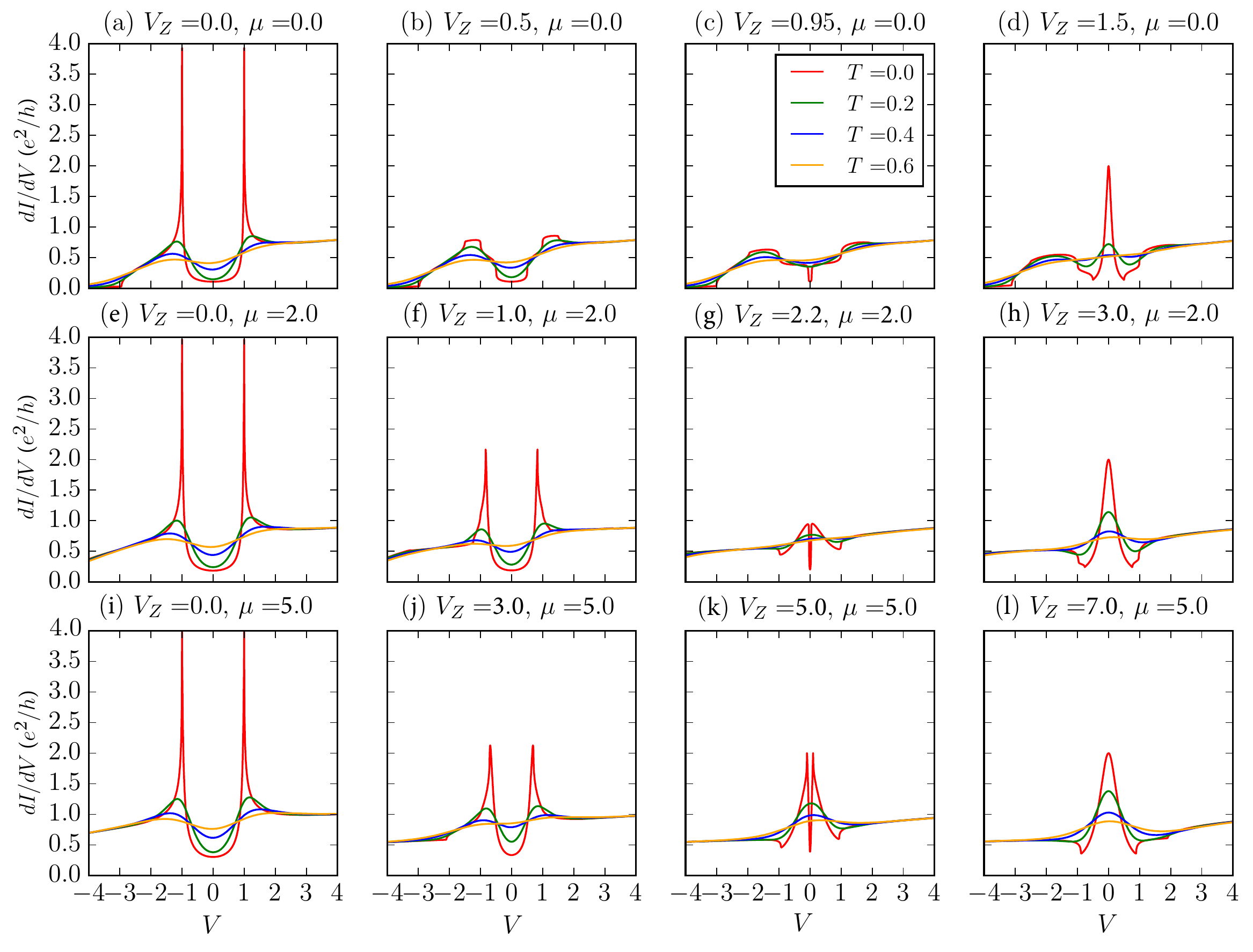}
\end{center}
\caption{(color online) Differential conductance $dI/dV$ vs $V$ for barrier strength $Z = 10.1$ and different temperature ($T$), Zeeman field ($V_Z$) and chemical potential of the nanowire ($\mu$). The Zeeman field increases from the left to the right panel where $V_Z$ for the second rightmost panel is just below the critical value of $V_Z$ where TQPT happens. For the rightmost panel, the system is in the topological phase where the ZBCP develops in the $dI/dV$ curve. The chemical potential $\mu$ increases from the top to the bottom panel.  The value of $\Delta$ is taken to be 1.}\label{fig:dIdV_diff_temp_nogrid_30000} 
\end{figure*}

\begin{figure*}[h]
\begin{center}
\includegraphics[width=\linewidth]{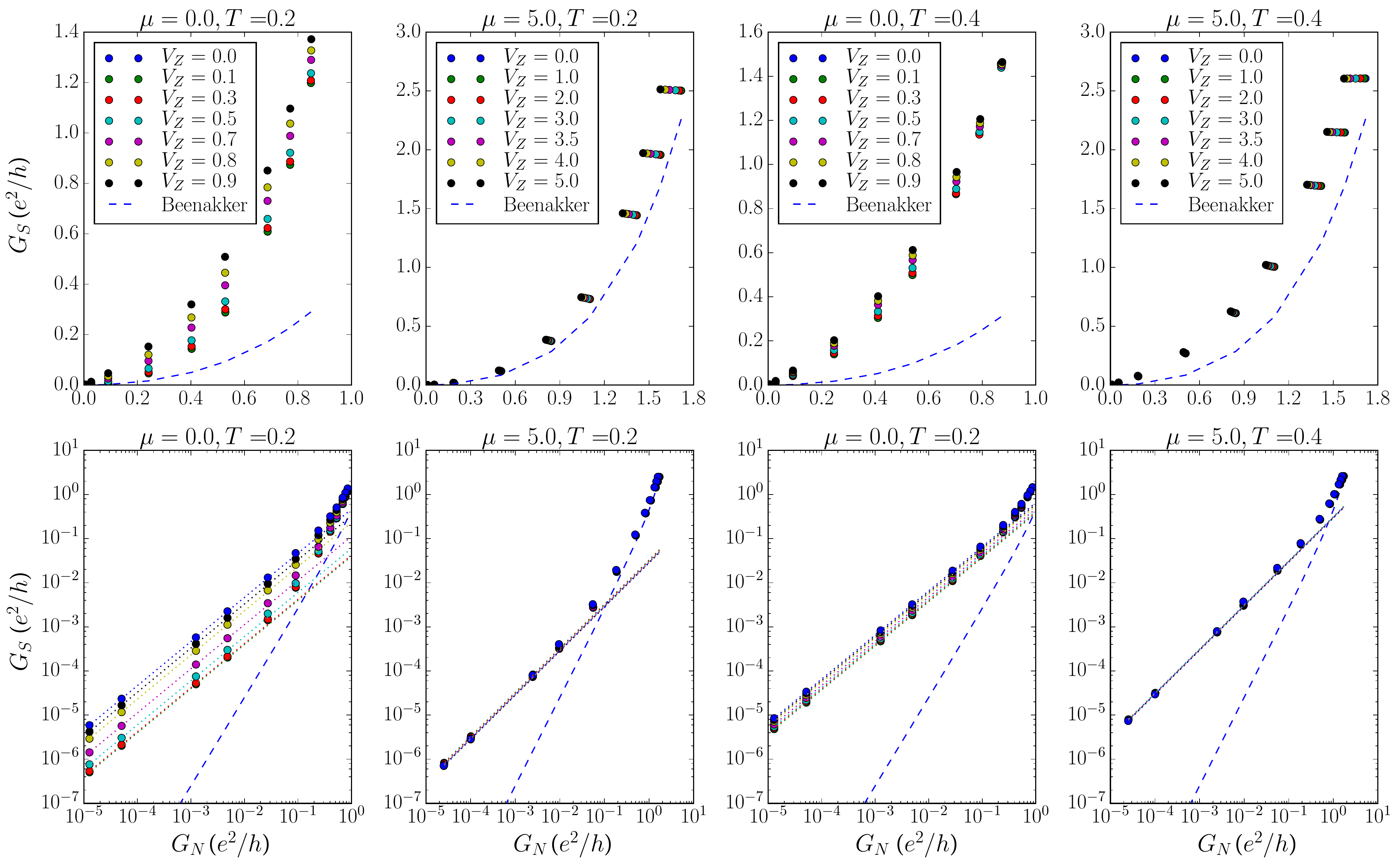}
\end{center}
\caption{(color online) Linear (top panel) and log-log plots (bottom panel) of finite-temperature ($T = 0.2$ and $T = 0.4$) values of $G_S$ vs $G_N$ for different chemical potential ($\mu = 0$ and $\mu = 5$) and Zeeman field $V_Z$ below the TQPT.  The value of $\Delta$ is taken to be 1. The dashed line is the Beenakker formula $G_{S} = (G_N)^2/(2 - G_N/2)^2$. Note that the Beenakker formula does not hold here for the finite-temperature case. The dotted line is the fit to the formula $G_S \sim G_N$. }\label{fig:GNGS_diff_mu_temp_02} 
\end{figure*}
\twocolumngrid

Figure~\ref{fig:dIdV_diff_temp_nogrid_30000} shows the evolution of the conductance as the Zeeman field is raised for different temperatures and chemical potentials. As can be seen from the figure, the coherence peak strength decreases as the Zeeman field increases. For low chemical potential and not very-high tunnel barrier strength, the superconducting gap may not appear as peaks in the conductance spectrum at finite Zeeman field [see Figs.~\ref{fig:dIdV_diff_temp_nogrid_30000}(b) and (c)].

As the Zeeman field is raised toward the TQPT, the gap shrinks. For high enough chemical potential and low temperature, near the TQPT, these two gaps appear as two coherence peaks near zero voltage. If the temperature is sufficiently high, these two coherence peaks can merge into one peak at zero voltage which resembles the ZBCP characteristic of the MZM (see panels (g) and (k) of Fig.~\ref{fig:dIdV_diff_temp}). When the Zeeman field is above the critical value, the ZBCP which signifies the presence of MZM appears. The ZBCP at zero temperature is universally quantized at $2e^2/h$ independent of the junction details. As the temperature increases, the ZBCP broadens and its peak value decreases. In fact, for $T$ high enough, the ZBCP could be arbitrarily small above the TQPT point, thus seriously compromising its topological properties (while also making the gap very soft because of thermal broadening). At $T=0$, by contrast, the ZBCP height is always $2e^2/h$ independent of the tunnel barrier strength $Z$.

To see how the subgap conductance changes at finite temperature, we plot the values of $G_S$ vs $G_N$ in Fig.~\ref{fig:GNGS_diff_mu_temp_02} for two different temperatures $T = 0.2$ and $T = 0.4$. In the tunneling limit, $G_S$ is related to $G_N$ through $G_S \sim G_N^2 + G_N \exp(-\Delta/T)$ where the first term arises from Andreev reflections and the second term from the thermal excitations of quasiparticles. In the limit of very large tunnel barrier (very small junction transparencies), the quasiparticle excitation process dominates and hence, the $G_S$ vs $G_N$ curve obeys the relation $G_S \sim G_N$ (as shown by the dotted lines in the bottom panel of Fig.~\ref{fig:GNGS_diff_mu_temp_02}). The range of $G_N$ over which this relation holds is $\sim \exp(-\Delta/T)$, which increases as $T$ increases (see the bottom panel of Fig.~\ref{fig:GNGS_diff_mu_temp_02}) because the thermal correction to the subgap conductance is exponentially weak for $\Delta \gg T$. For higher $T$, the thermal correction to $G_S$ is considerable and $G_S \sim G_N$ holds producing a very soft gap even in the weak-tunneling limit. In the regime where $T > \Delta$, braiding with MZMs cannot be performed successfully because of the thermal excitations of the quasiparticles. For the value of $G_N$ above this range but still in the tunneling limit, $G_S \sim G_N^2$ due to Andreev reflection. Note that for the finite-temperature case, $G_S$ vs $G_N$ curve deviates from the Beenakker formula even for $V_Z = 0$ and large $\mu$.

We mention that the semiinfinite BTK model used in this section does not allow any MZM overlap since the nanowire is by definition infinitely long, and thus the physics of MZM oscillation does not appear in this model even for very large $V_Z$ values. We consider finite nanowires in the next section where we also include the effect of dissipative broadening in the theory (left out in the results shown in the current section) to better simulate realistic systems.


\section{Finite nanowires}\label{sec:realistic}

In this section, we numerically calculate the differential conductance ($G=dI/dV$) of finite-length NS junctions as a function of bias voltage at increasing Zeeman field studying its dependence on various intrinsic parameters, e.g., strength of the tunnel barrier $Z$, temperature $T$, dissipation $\Gamma$, and length of the nanowire $L$. The dissipation is phenomenologically modeled as $i\Gamma$ in the BdG Hamiltonian of the nanowire (Eq.\eqref{eq:H_NW}), the origin of which is speculated to be vortices or other quasiparticle states in the parent superconductor. For the details on how dissipation is introduced into the theory, we refer to Refs.~\cite{DasSarma2016How, Liu2017Role}--all one does is to add an $i\Gamma$ term to Eq.\eqref{eq:H_NW} before calculating the conductance.  Similar to the discussion of semiinfinite nanowires in the previous section, the focus here is on understanding how a soft gap develops in the nanowire and how it evolves with increasing Zeeman field, eventually going through the TQPT and forming ZBCP. In the absence of finite temperature and dissipation, the ZBCP should always have the quantized value $2e^2/h$ independent of how soft the induced gap might be because of strong-tunneling effect. An equally important thing we are trying to understand is how the ZBCP quality including its height and width changes with respect to these intrinsic parameters. Furthermore, we study how the splitting of the ZBCP is affected by various intrinsic parameters, which is a property unique to finite-length nanowires since the splitting of the ZBCP is due to the overlap between the two MZMs at the end of the nanowire, i.e., ZBCP splits because of MZM splitting~\cite{Cheng2009Splitting, Cheng2010Tunneling, DasSarma2012Splitting}.

\subsection{Effect of tunnel barrier}\label{subsec:barrier}
\begin{figure}
\includegraphics[width=0.47\textwidth]{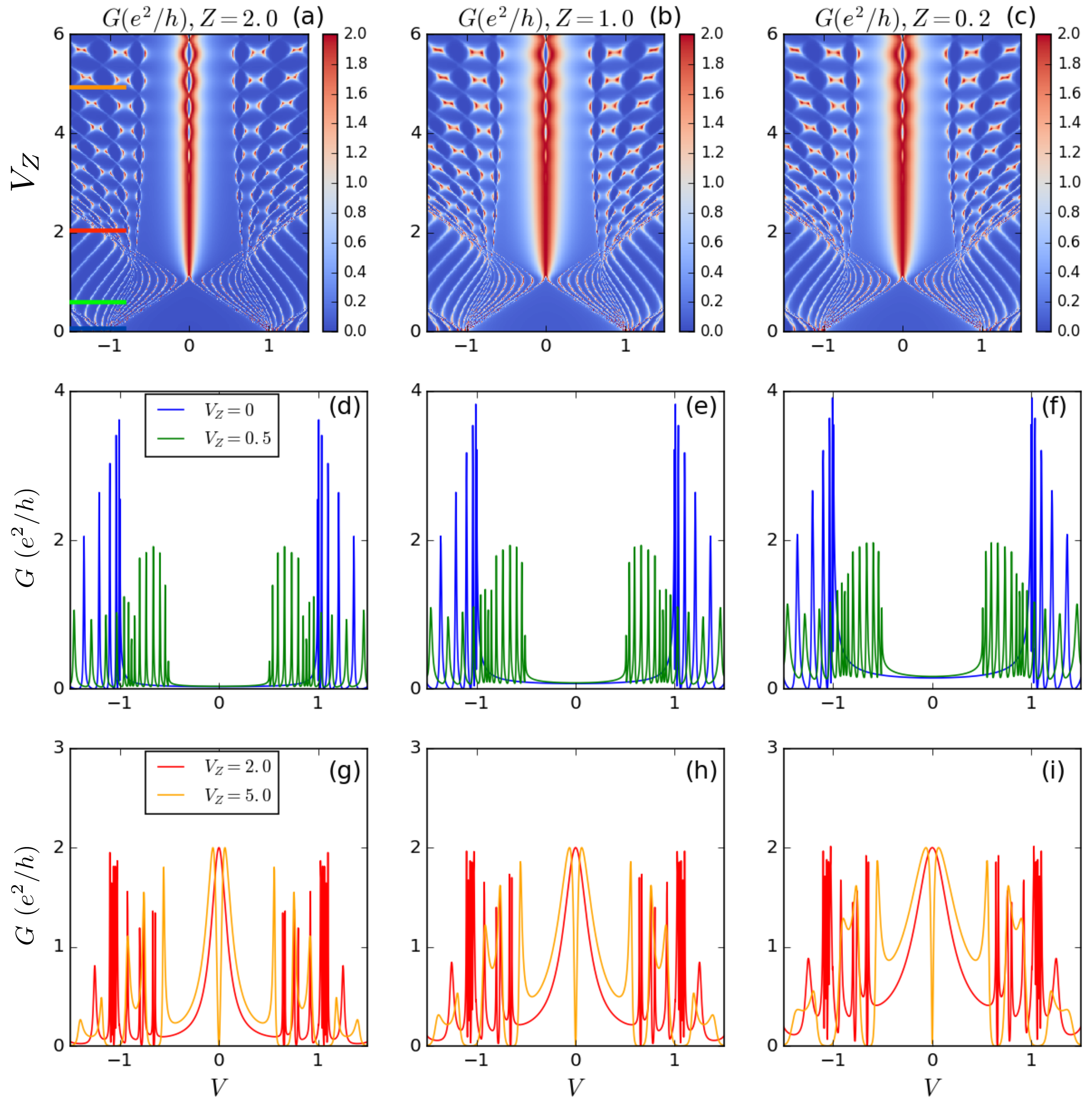}
\caption{(color online) Conductances of NS junction with various barrier strength $Z$ at zero temperature, zero dissipation. In the upper panels, we show the conductance as a function of bias voltage $V$ at increasing Zeeman spin splitting $V_Z$, and the two lower panels are the conductance linecuts of the corresponding color plots in the upper panels. The barrier strength $Z$ from the left to the right panels are $Z=2.0, 1.0, 0.2$. The parameters used in all the panels are $L=1.3\mu$m, $\mu=0, \Delta=1$ and thus, $V_{Zc}=1$. }
\label{fig:realbarrier}
\end{figure}

The tunnel barrier at the NS junction between the normal lead and the finite-length superconducting nanowire is modeled as a delta-function barrier with strength $Z$ as in Sec.~\ref{sec:semiinfinite}. In order to study the effect of tunnel barrier on the differential conductance, we vary the barrier strength $Z$ while keeping both temperature and dissipation to be precisely zero and fixing the nanowire length, for which the numerical results are shown in Fig.~\ref{fig:realbarrier}. The upper panels show the conductance as a function of bias voltage $V$ at increasing Zeeman splitting $V_Z$, and the two lower panels show the conductance linecuts of the color plots in the upper panels. From the left to the right panels, the barrier strength is lowered. The conductance plots in all the upper panels show the generic trend that there is an induced gap at zero Zeeman field between the coherence peaks. The gap shrinks with increasing Zeeman field until the Zeeman field crosses the critical value $V_{Zc}$, where the gap closes and reopens, indicating TQPT. Above the critical Zeeman field, a ZBCP forms with a peak value $2e^2/h$. When the Zeeman field is even larger, the ZBCP splits, and the amplitude of the splitting oscillates with the Zeeman field, which is the Majorana oscillation due to the overlap between two MZMs at the wire ends~\cite{Cheng2009Splitting, Cheng2010Tunneling, DasSarma2012Splitting}. Our first observation is that in the non-topological regime ($V_Z<V_{Zc}=1$), the induced gap can be softened by lowering the barrier strength (as shown in Figs.~\ref{fig:realbarrier}(d)-(f)). From Fig.~\ref{fig:realbarrier}(d) to (f) the conductance at zero-bias voltage increases from $\sim 0.03$ to $\sim 0.15e^2/h$ when the barrier strength $Z$ decreases from 2 to 0.2. As $V_Z$ increases the induced gap shrinks (as shown by the green lines). Second, in the topological regime ($V_Z > 1$), the red linecuts in Figs.~\ref{fig:realbarrier}(g)-(i) show how the quality of MZM-induced ZBCP changes with the tunnel barrier. With the decrease of the barrier strength, the peak value of the ZBCP stays at $2e^2/h$ but the ZBCP width increases, covering a larger portion of the topological gap, making the topological gap soft. Third, at even larger Zeeman field way above TQPT, yellow linecuts show the feature of Majorana oscillations of the ZBCP, where the ZBCP splitting is characterized by a dip at the zero-bias voltage. Note that not only is the conductance at zero-bias voltage always zero, also the separation of the split peaks is constant--both quantities are independent of the barrier strength. Strictly speaking, for a finite wire at zero temperature and zero dissipation, the conductance precisely at zero bias must always be zero way above the TQPT since there is always a little MZM energy splitting in any finite wire and single-channel approximation becomes valid~\cite{DasSarma2016How}. No amount of tunneling by itself can produce finite conductance at zero bias in the topological regime way above TQPT. On the other hand, the yellow linecuts show that the width of the split ZBCP also increases when the barrier strength is lowered. The broadened ZBCP will cover the whole topological gap in the large Zeeman field regime.

To sum up, the effect of tunnel barrier on the differential conductance is at least three-fold. First, in the non-topological regime, the decrease of the barrier strength will soften the gap by increasing the Andreev conductance inside the induced gap. Second, in the topological regime where Zeeman field is slightly greater than the critical field, lowering the barrier strength broadens the width of the ZBCP while keeping the height unchanged. The ZBCP, with the broadening effect, will cover a larger region inside the topological gap, making the topological gap softer. Third, at even larger Zeeman field, the ZBCP splits, i.e., the Majorana oscillation shows up, but the splitting is not affected by the barrier strength, although the ZBCP gets broadened, covering the whole topological gap. We emphasize, however, that the ZBCP remains quantized at $2e^2/h$ no matter how soft the gap might be as long as the softness is caused purely by tunneling effects.

\subsection{Effect of temperature}\label{subsec:temperature}
\begin{figure}
\includegraphics[width=0.47\textwidth]{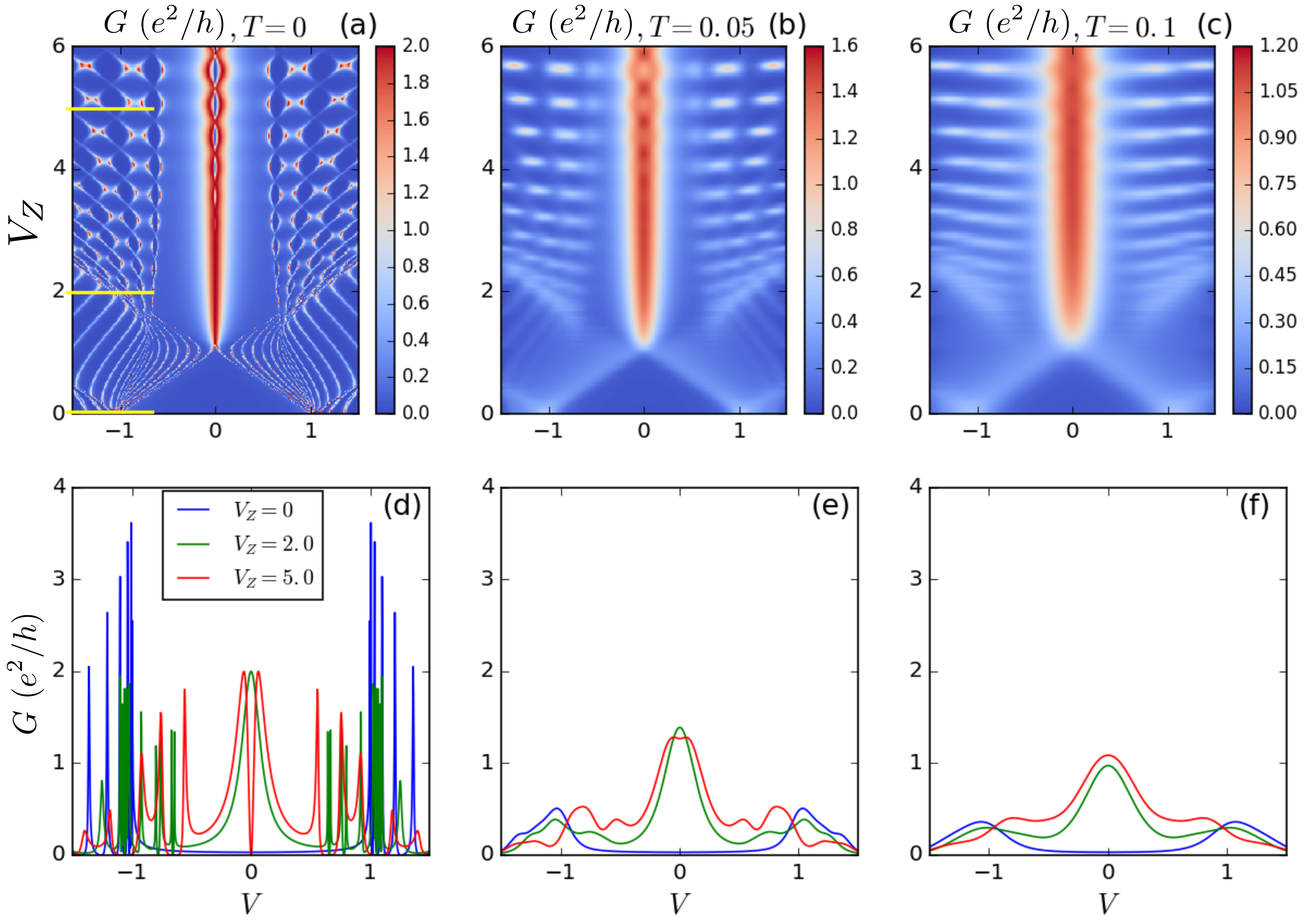}
\caption{(color online) Conductances of NS junction at various temperature with fixed barrier strength and zero dissipation. In the upper panels [(a)-(c)] we show the conductance color plots as a function of the bias voltage at increasing Zeeman field. The lower panels [(d)-(f)] are the conductance linecuts of the color plots at various fixed Zeeman fields. The temperatures in the left, middle and right panels are $T=0, 0.05, 0.1$, respectively. The parameters used in all panels are $L=1.3\mu$m, $Z=2.0$, $\mu=0, \Delta=1$ and thus, $V_{Zc}=1$.}
\label{fig:realtemperature}
\end{figure}

Temperature is an essential ingredient in the tunneling experiment of NS junctions. Although the base temperature is known as the fridge temperature in the experiment, the electron temperature in the nanowire might be much higher than the base temperature due to lack of thermalization. This is in fact a well-known problem in low-dimensional semiconductor structures. Thus, temperature is indeed an unknown parameter in the experimental setup. We, therefore, theoretically study the temperature dependence of the conductance profile at fixed tunnel barrier ($Z=2.0$) and zero dissipation ($\Gamma=0$). As shown in Eq.~\eqref{eq:GT}, finite temperature effect is introduced by a convolution between the zero-temperature conductance and the derivative of the Fermi-Dirac distribution. The numerically calculated conductance at various temperature is shown in Fig.~\ref{fig:realtemperature}. In the upper panels, the differential conductance is plotted as a function of the bias voltage at increasing Zeeman field. The lower panels are the conductance linecuts at various fixed Zeeman fields. From the left to the right panels, the temperature increases. The generic feature of the conductance color plot is that the induced gap shrinks with increasing Zeeman field, and the gap closes and reopens with the formation of the topological ZBCP when the Zeeman field crosses the critical value. First we will try to understand how the induced gap is softened by increasing temperature. As we can see from Figs.~\ref{fig:realtemperature}(d)-(f), the blue lines are the linecuts of the conductance at zero Zeeman field with high tunnel barrier, when the temperature is increased, the initially hard induced gap is softened in the sense that the conductance below the gap edge within the magnitude $\sim T$ becomes finite, i.e. $G(V) > 0$ when $\Delta - T \lesssim |V| < \Delta$, although the conductance in the middle of the induced gap is still negligibly small, i.e. $G(V) \simeq 0$ when $|V| \ll \Delta$. This is because the temperature convolution is effectively an average of conductance with its neighboring values within an energy range $\delta V \simeq T$. Thus, if the initial induced gap is very hard, only the conductance close to the induced gap edge by $\sim T$ will be enhanced due to temperature. In the topological regime [as shown by the green lines in Figs.~\ref{fig:realtemperature}(d)-(f)], the MZM-induced ZBCP gets broadened and its peak value gets lowered simultaneously, but the total area ($\sim$ height $\times$ width) remains almost the same~\cite{Liu2017Role}. The broadened ZBCP will cover a larger region inside the topological gap, making the topological gap soft. Such a behavior of the quality of ZBCP with respect to rising temperature is in sharp contrast with the effect due to lowering the barrier strength, where the ZBCP gets broadened but its peak value remains unchanged. When the Zeeman field is even larger, the ZBCP splitting can be observed at least at zero temperature, as shown by the red lines in Fig.~\ref{fig:realtemperature}(d). But when the temperature rises, the ZBCP splitting is smeared due to thermal averaging and thus the Majorana oscillations disappear along with the broadening of ZBCP with its height going down. This is also a stark difference from the decreasing barrier situation, where the ZBCP splitting remains exactly the same no matter what value the barrier strength is. It is clear that finite temperature (or dissipation, see below) is essential for suppressing MZM oscillations--tunnel barrier physics by itself cannot eliminate ZBCP oscillations. Note also that finite temperature, by eventually strongly suppressing the ZBCP value, suppresses the Majorana topological behavior~\cite{DasSarma2016How}.

\subsection{Effect of dissipation}\label{subsec:dissipation}
\begin{figure}
\includegraphics[width=0.47\textwidth]{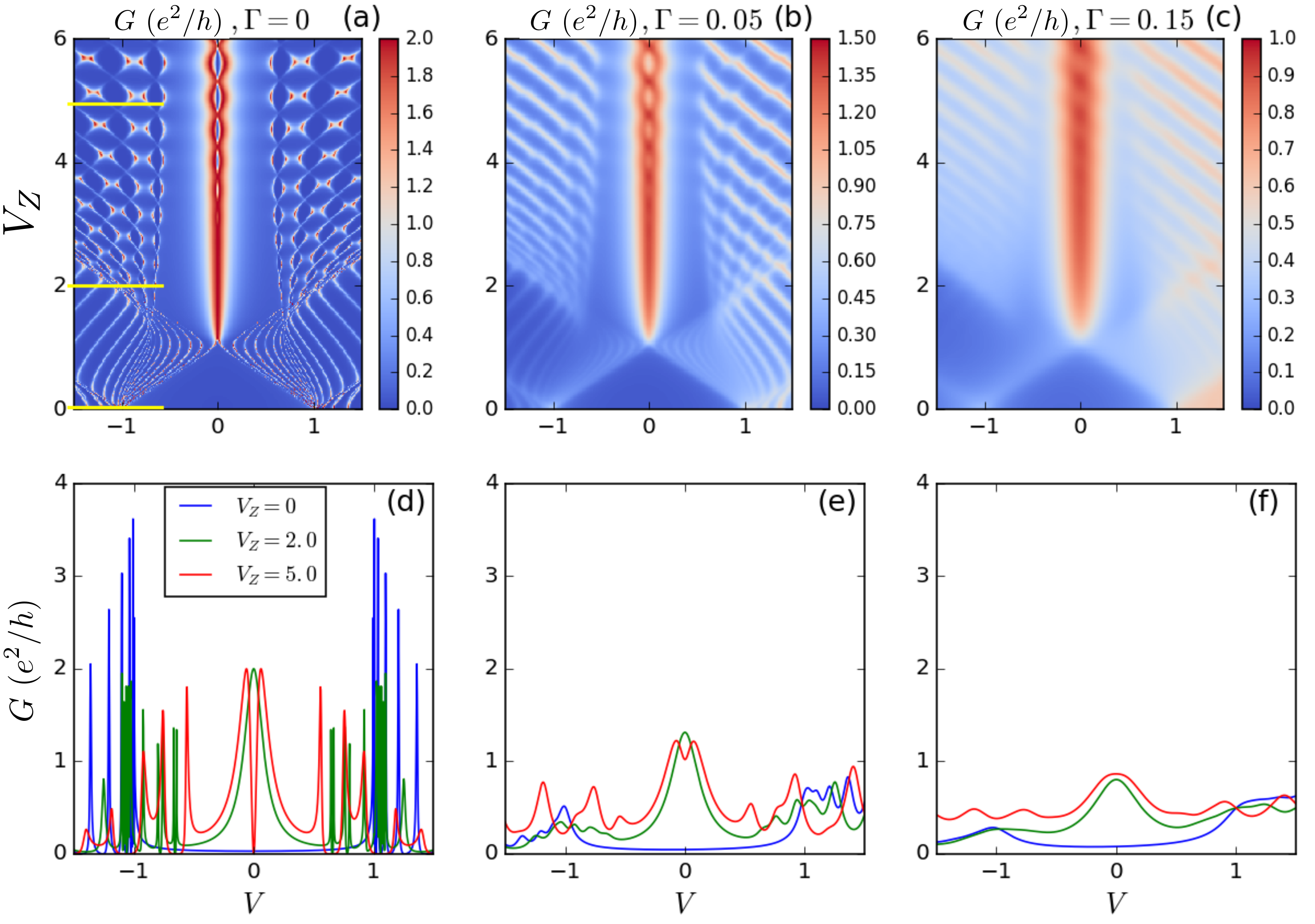}
\caption{(color online) Conductances of NS junction with various amount of dissipation but with fixed barrier strength and zero temperature. In the upper panels [(a)-(c)] are the differential conductances plotted as a function of the bias voltage at increasing Zeeman field. The lower panels [(d)-(f)] are the conductance linecuts at various fixed Zeeman fields. The dissipation strengths in the left, middle and right panels are $\Gamma=0, 0.05, 0.15$, respectively. The parameters used in all the panels are $L=1.3\mu$m, $Z=2.0$, $\mu=0$, $\Delta=1$ and thus, $V_{Zc}=1$.}
\label{fig:realdissipation}
\end{figure}

Dissipation is another essential ingredient for understanding the tunneling experiment of NS junctions~\cite{DasSarma2016How, Liu2017Role}. Physically, there probably exist vortices or quasiparticle states in the parent superconductor coupling with the nanowire, leading to dissipation in the nanowire. Theoretically, we simply model all dissipative mechanisms by a single phenomenological parameter $\Gamma$, which simulates coupling of the superconducting nanowire to any fermionic bath. In order to see the $\Gamma$-dependence of the differential conductance, we numerically calculate the conductance of an NS junction at zero temperature with fixed tunnel barrier. The focus is on how the gap softening, the quality of the MZM-induced ZBCP, and the Majorana oscillations are affected by dissipation. Numerical results are summarized in Fig.~\ref{fig:realdissipation}. In the upper panels of Fig.~\ref{fig:realdissipation}, we show the differential conductance as a function of the bias voltage with increasing Zeeman field. The lower panels are the conductance linecuts at various fixed Zeeman fields. From the left to the right panels, the dissipation increases. The blue lines in Figs.~\ref{fig:realdissipation}(d)-(f) are the linecuts of the conductance at zero Zeeman field. When the dissipation increases, the hard induced gap is softened, i.e., the conductance inside the gap increases everywhere. In the topological regime, as shown by the green lines in Figs.~\ref{fig:realdissipation}(d)-(f), the MZM-induced ZBCP gets broadened and its height goes down simultaneously, with the total area ($\sim$ height $\times$ width) remaining almost the same~\cite{Liu2017Role}. At even larger Zeeman field, the ZBCP splitting, which can be observed at zero dissipation, gets smeared and disappears with increasing dissipation (as shown by the red lines in the lower panels of Fig.~\ref{fig:realdissipation}). On the whole, dissipation in the topological regime will make the ZBCP broaden and cover more region inside the topological gap, making the gap soft. Furthermore, a unique effect of dissipation, is that it introduces particle-hole asymmetry into the outside-gap conductance profile~\cite{Liu2017Role} (as indicated by Figs.~\ref{fig:realdissipation}(b), (c), (e) and (f)). To be precise, the differential conductances at positive and negative bias voltages are different from each other. Based on all these observations, we see that all the effects that the dissipation has on the conductance except for the particle-hole asymmetry, show strong similarities with those of finite temperature, e.g., gap softening, broadening of ZBCP, height of ZBCP going down, smearing of Majorana oscillations. By contrast, the effect of barrier strength (Sec.~\ref{subsec:barrier}) is qualitatively different as it only introduces gap softening. Therefore, in the next subsection we will make a direct comparison between the effects of dissipation and finite temperature.

\subsection{Direct comparison between effects of temperature and dissipation}\label{comparing}
\begin{figure}
\includegraphics[width=0.47\textwidth]{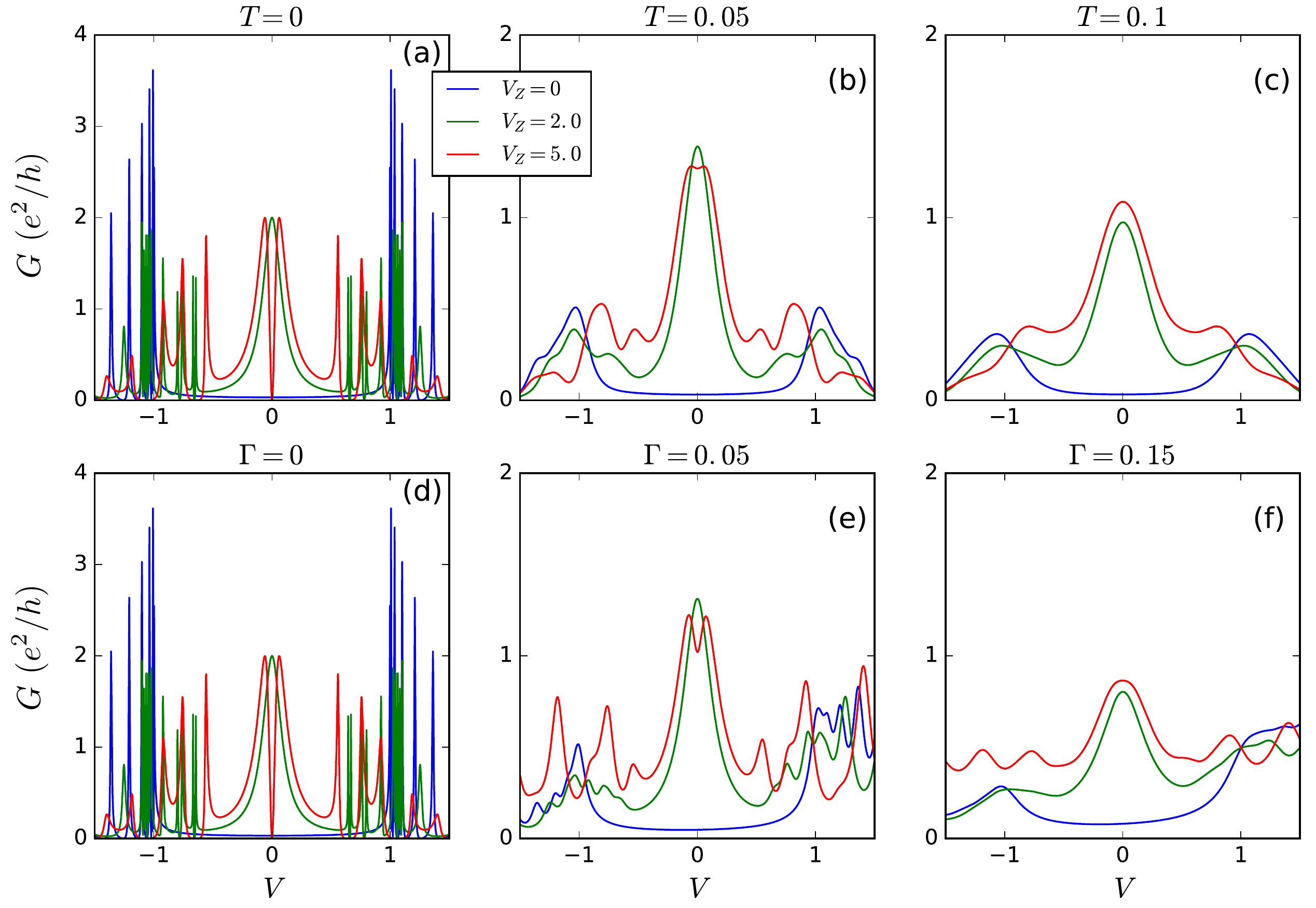}
\caption{(color online) Conductance plots at various temperatures or dissipations. In panels (a)-(c) are the conductances at increasing temperature with no dissipation in the system, while in panels (d)-(f) are the conductances for systems with increasing amount of dissipation but at zero temperature. The parameters used in all panels are $L=1.3\mu$m, $Z=2.0$, $\mu=0, \Delta=1$ and thus, $V_{Zc}=1$.}
\label{fig:comparison}
\end{figure}

Based on the discussion of the effects of temperature and dissipation on the differential conductance in previous subsections, we can see quite a few similarities and some differences. Thus, this subsection is devoted to a direct comparison between the temperature and dissipation effects on the differential conductance. Numerical results are shown in Fig.~\ref{fig:comparison} for finite-length nanowire with fixed tunnel barrier strength. In the upper panels, we show the conductance at various temperature with no dissipation in the system, while in the lower panels we show the conductance for systems with finite dissipation but at zero temperature. We see that at zero Zeeman field (blue lines), both raising temperature and increasing dissipation enhance the conductance inside the induced gap, softening the gap in a similar manner. Above the TQPT (green lines), either increasing the temperature or dissipation will broaden the ZBCP width and lower its height simultaneously, while keeping the ZBCP area ($\sim$ height $\times$ width) approximately the same. At even larger Zeeman field (red lines), although the ZBCP splitting is clearly observed at zero temperature or zero dissipation, the splitting is smeared away by finite temperature or dissipation effect. Meanwhile, the conductance at zero-bias voltage is also no longer zero. The tendency of smearing away the ZBCP splitting and giving finite value to zero-bias conductance is shared by both finite temperature and dissipation effect, but is not observed by lowering the barrier strength. It indicates that if the ZBCP splitting or zero conductance at zero bias is not observed in experiments, it is evidence of finite temperature or existence of dissipation in the NS junction system. Apart from similar effects on gap softening, quality of ZBCP and Majorana oscillations, a qualitative difference between temperature and dissipation effects is that dissipation breaks the particle-hole symmetry of the differential conductance while temperature does not (as shown in Fig.~\ref{fig:comparison}). Note that only conductance at finite bias voltage shows particle-hole asymmetry with dissipation, the ZBCP is always particle-hole symmetric~\cite{Liu2017Role}. This observation indicates that any degree of particle-hole asymmetry in the experimental data could be a clue to the existence of dissipation inside the system. Other than the particle-hole asymmetry aspect, temperature and dissipation produce similar observable effects on the differential conductance.

\subsection{Effect of increasing Zeeman field}\label{subsec:Vz}

\begin{figure}[h]
\includegraphics[width=0.47\textwidth]{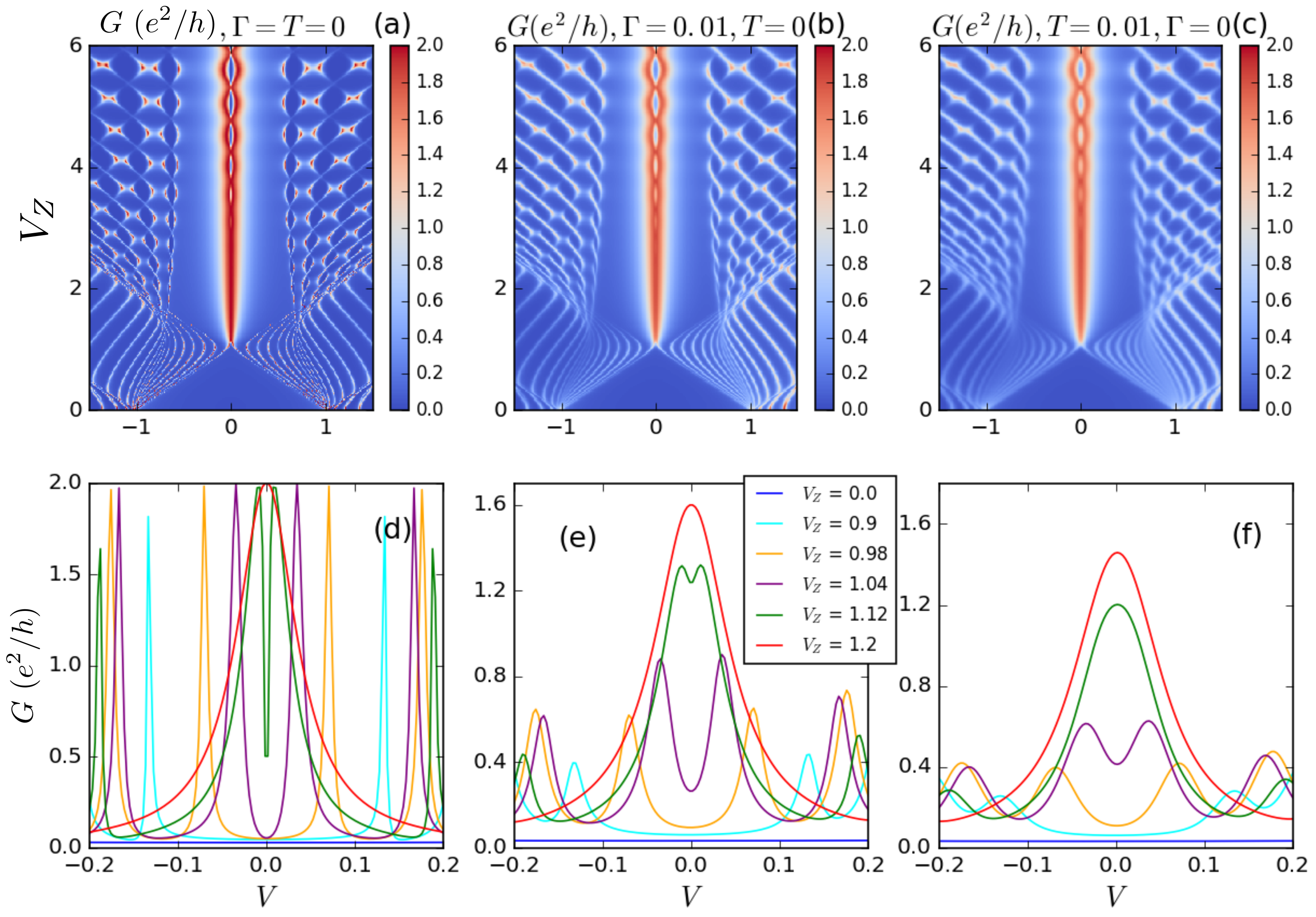}
\caption{(color online) In panels (a)-(c) are the conductance color plots as a function of bias voltage $V$ at increasing Zeeman field $V_Z$, and in panels (d)-(f) are the corresponding conductance linecuts at various fixed Zeeman fields. The left panels [(a) and (d)] are conductances at zero temperature and with zero dissipation. The middle panels [(b) and (e)] are conductances with dissipation $\Gamma=0.01$ at zero temperature, and the right panels [(c) and (f)] are conductances at finite temperature $T=0.01$ with no dissipation. The parameters used in all panels are $L=1.3\mu$m, $Z=2.0$, $\mu=0, \Delta=1$ and thus, $V_{Zc}=1$.}
\label{fig:Vz}
\end{figure}

In this subsection, we study the effect of increasing Zeeman field on the differential conductance, especially focusing on the effect of Zeeman field on the conductance below TQPT and on MZM splitting. The numerical calculation of the conductance of a finite-length nanowire ($L=1.3\mu$m) is shown in Fig.~\ref{fig:Vz}. In the upper panels, the conductance is plotted as a function of bias voltage $V$ at increasing Zeeman field $V_Z$, while the lower panels are conductance linecuts at various fixed Zeeman fields from the corresponding color plots in the upper panels. The conductance in the left panels is at zero temperature and with no dissipation, while the conductance in the middle and right panels is at finite dissipation and finite temperature, respectively. Figures~\ref{fig:Vz}(d), (e) and (f) show that the subgap conductance below TQPT increases with increasing Zeeman field. This enhancement of subgap conductance is related to the fact that the coherence peaks get closer to each other. Such a trend of increasing conductance inside the gap is also consistent with results in Sec.~\ref{sec:semiinfinite}. On the other hand, when the Zeeman field is much larger than the critical Zeeman field for TQPT, splitting of the ZBCP can be observed due to the overlap between two MZMs at the wire ends. The amplitude of the ZBCP splitting oscillates with increasing Zeeman field, which is the manifestation of Majorana oscillations~\cite{Cheng2009Splitting, DasSarma2012Splitting} (as shown in Figs.~\ref{fig:Vz}(a), (b) and (c)). On the whole, the amplitude of the Majorana oscillation increases with the Zeeman field, since in the large Zeeman limit, the effective superconducting gap decreases with Zeeman field, making the superconducting coherence length longer~\cite{DasSarma2012Splitting} and consequently, the effective wire length shortens.

\subsection{Effect of the nanowire length}\label{subsec:long}
\begin{figure}
\includegraphics[width=0.47\textwidth]{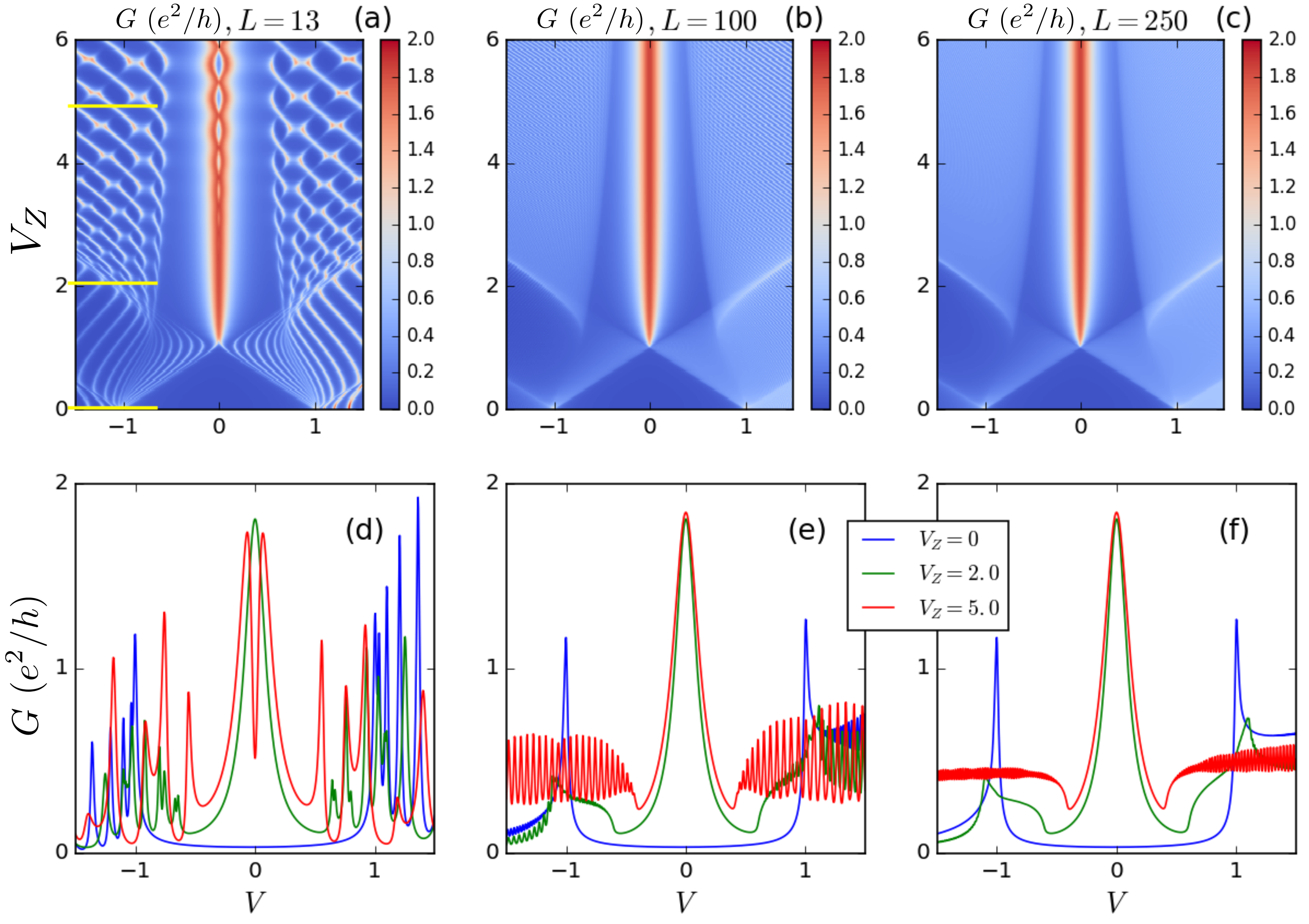}
\caption{(color online) Conductances of NS junction with various wire lengths with fixed tunnel barrier strength $Z=2.0$, dissipation $\Gamma=0.01$ and at zero temperature $T=0$. In the upper panels [(a)-(c)], we show conductance as a function of bias voltage $V$ at increasing Zeeman field $V_Z$, while the lower panels [(d)-(f)] are conductance linecuts at various fixed Zeeman field from the corresponding upper panel. The nanowire lengths in the left, middle and right panels are $L=13, 100, 250 $ in units of $l_{SO}=0.1\mu$m. The parameters used in all panels are $Z=2.0$, $\mu=0$, $\Delta=1$ and thus, $V_{Zc}=1$.}
\label{fig:reallong}
\end{figure}

\begin{figure}
\includegraphics[width=0.47\textwidth]{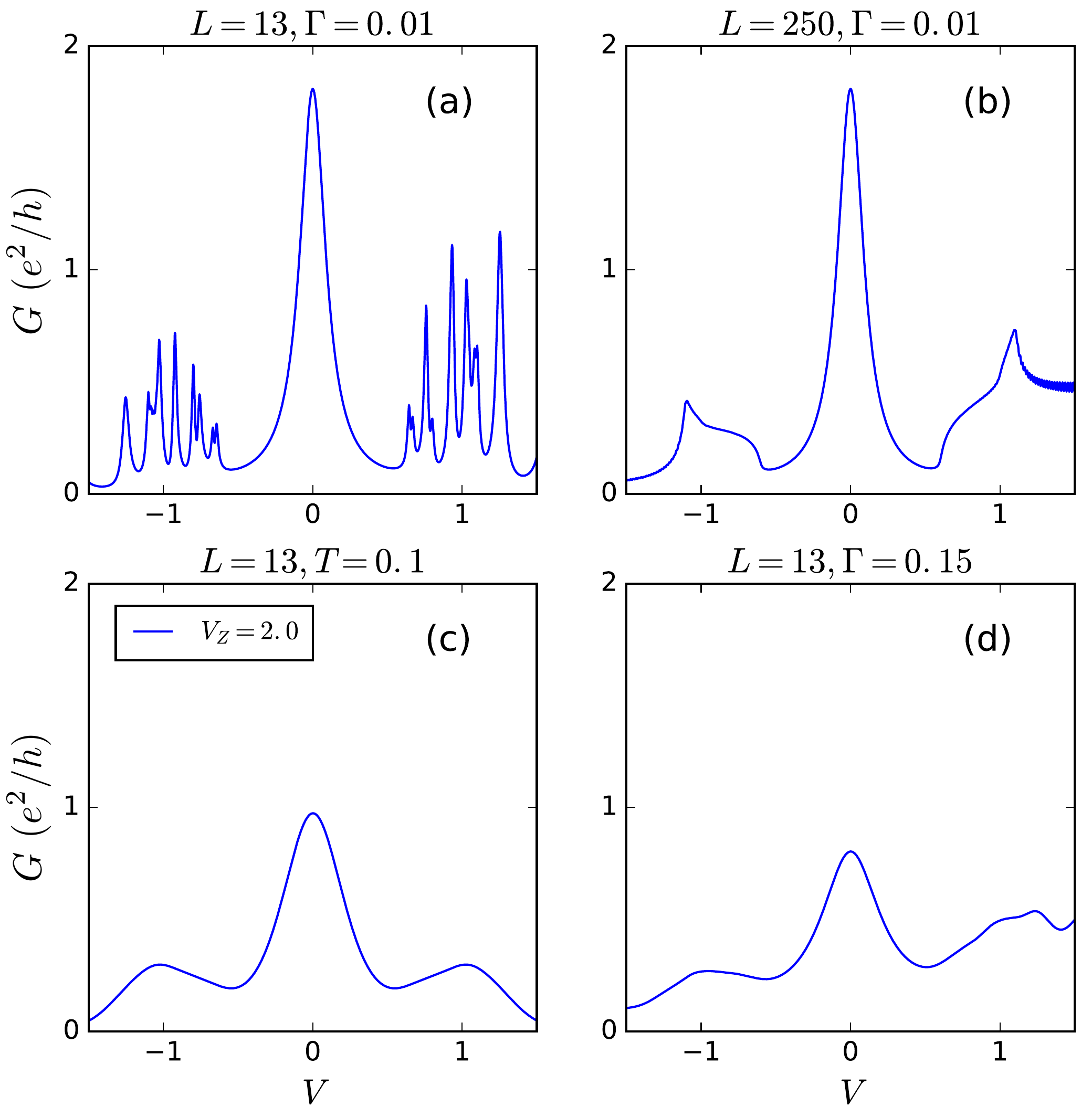}
\caption{(color online) Smoothness of the conductance due to different mechanisms. (a) Conductance of a short nanowire $L=13$ with small dissipation $\Gamma=0.01$. (b) Conductance for a long nanowire $L=250$ with small dissipation $\Gamma=0.01$. (c) Conductance for a short nanowire $L=13$ at finite temperature $T=0.1$ but without dissipation. (d) Conductance for a short nanowire $L=13$ with finite dissipation $\Gamma=0.15$. The nanowire length is given in units of $l_{SO}=0.1$ $\mu$m. The parameters used in all panels are $Z=2.0$, $\mu=0, \Delta=1$ and thus, $V_{Zc}=1$. }
\label{fig:reallong2}
\end{figure}

In the previous subsections, we discuss the differential conductance for NS junctions with finite-length nanowires. The focus is on how the gap softening, the quality of ZBCP including its height and broadening, and the Majorana oscillations are affected by various intrinsic parameters, e.g. tunnel barrier, temperature and dissipation. All of these are subgap features of conductance, whether the gap is the induced gap in the non-topological regime or topological gap in the topological regime. On the other hand, however, we can hardly define any meaningful quantities like normal conductance $G_N$ outside the gap for finite-length wires in previous subsections, based on the numerical simulations shown in Figs.~\ref{fig:realbarrier}-\ref{fig:realdissipation}, in contrast with semiinfinite nanowire situations (Sec.~\ref{sec:semiinfinite}), where $G_N$ is well defined. The reason is that for finite-length nanowire, the energy levels are discrete with the characteristic level spacing $\delta E \simeq 2\pi v_F/L$. Conductance, therefore, shows up as a set of coherence peaks separated by level spacing outside the gap. This makes outside-the-gap conductance $G_N$ look extremely noisy as these finite-size-induced nanowire energy levels all contribute to the above-gap conductance. In order to extract $G_N$ from plots and make a connection between the numerical results of finite-length nanowires and those of semiinfinite nanowires, we here numerically calculate the differential conductance for nanowires of increasing length with fixed barrier strength ($Z=2.0$) and at zero temperature ($T=0$), as shown in Fig.~\ref{fig:reallong}. Note that we need to add an infinitesimal amount of dissipation ($\Gamma=0.01$) into the system in order to smoothen the conductance curve. Our argument is that when the level spacing is less than the dissipation strength, i.e. $\delta E \simeq 2\pi v_F/L \lesssim \Gamma$, the conductance behavior will cross over to the conductance of semiinfinite nanowires. (Temperature will also act as a similar cut-off parameter for suppressing the finite-size noise outside the gap.) Oscillations are likely suppressed in the experiment because of the participation of several bands to above-gap conductance. The mechanism we consider (i.e., longer length together with small dissipation) is likely qualitatively similar to this effect. In the upper panels of Fig.~\ref{fig:reallong}, we show the calculated conductance as a function of bias voltage $V$ at increasing Zeeman field $V_Z$, while the lower panels are conductance linecuts at various fixed Zeeman field from the corresponding upper panels. The nanowire lengths in the left, middle and right panels are $L=13, 100, 250$ in units of spin-orbit length $l_{SO} \simeq 0.1 \mu$m. At zero Zeeman field, as shown by the blue lines in Figs.~\ref{fig:reallong}(d)-(f), the hard-gap feature ($G(V) \ll e^2/h$ for $-\Delta < V < \Delta$) is not affected by the length of the nanowire at all, while the conductance peak at induced gap becomes sharper. Above the TQPT (as shown by the green lines in panels (d)-(f)), the quality of the ZBCP, e.g. its height and width, is also not affected by the change of wire length. At even larger Zeeman field (as shown by the red lines in panels (d)-(f)), the splitting of the ZBCP at small length disappears when the length of the wire increases, since the ZBCP splitting exponentially decays with the wire length. For all of the conductance linecuts at various fixed Zeeman fields, the normal conductance outside the gap is smoothened by increasing the wire length, making normal conductance $G_N$ gradually well defined with the suppression of the noisy features. Especially for the longest nanowire in Fig.~\ref{fig:reallong}(f), the conductance crosses over to that for the semiinfinite nanowire shown in Fig.~\ref{fig:dIdV_VZ_barrier}. On the other hand, based on the discussions in previous subsections, the smoothness of the conductance linecuts is not unique to long nanowires, smoothness can also show up when the system contains dissipation or is at finite temperature (see Figs.~\ref{fig:reallong2}(c) and (d)).

Figure~\ref{fig:reallong2} shows that the above-gap conductance becomes smooth as the length of the nanowire, temperature and dissipation increase. The main difference in the conductance profiles is that finite temperature or dissipation would lower the height of ZBCP significantly while the longer nanowire does not. Since in the experiment, all the conductance profiles are smooth~\cite{Deng2016Majorana, Gul2017Hard, Zhang2017unknown}, it is highly probable that either the system contains dissipation, or the effective temperature is high or the effective length of the nanowire is long. Since the typical nanowire length in the experiment is not long ($L \sim 1$ $\mu$m) and the temperature is generally low ($T \sim 100$ mK) we conclude that the current experiment systems may have considerable dissipation, explaining the absence of size-quantized level structure in the above-gap conductance.

To sum up, most of the in-gap features, e.g. the hard gap, the quality of ZBCP, etc. are unaffected by the length of the nanowire, except for the Majorana oscillations, which disappear for long nanowires. Outside the induced gap, however, the conductance becomes smoother and the normal conductance $G_N$ is better defined when wire length increases. This smoothness of above-gap conductance is similar to the effects of dissipation and finite temperature.


\section{Discussion and conclusion}\label{sec:discussion}

Using a minimal theoretical model (to keep unwanted complications to a minimum),  we have systematically studied effects of a number of intrinsic physical parameters on the phenomenology of the Majorana nanowire physics in superconductor-semiconductor hybrid structures, focusing specifically on the issue of the soft induced gap where subgap differential conductance is often found to be rather large ($\sim e^2/h$) instead of being exponentially small as expected for bulk superconductors. The subgap conductance depends on the tunnel barrier at the junction between the normal metal and the nanowire, and can be large even at zero temperature if the junction is transparent (i.e., the barrier low).  We have systematically studied the dependence of the gap softness on the junction transparency (parametrized by $Z$ in our theory and simulation), establishing that a very large subgap conductance (or equivalently a very soft gap) may arise intrinsically in the tunneling spectroscopy of the nanowire with no detrimental effect on the nanowire ZBCP quantization topological properties whatsoever. Our results providing the subgap conductance $G_S$ as a function of the barrier strength and/or the above-gap conductance $G_N$ (with $G_N$ being, by definition, an operational quantitative measure of the junction transparency) show that $G_S$ could be as large as $G_N$ itself when the junction barrier is very low. Thus, the gap could indeed appear to be very soft in tunneling measurements with \emph{no implications whatsoever for the topological nature of the system}, provided, of course, the gap softness is arising entirely from the large-junction transparency physics. The question,therefore arises as how one can ensure that the experimentally observed soft-gap behavior arises entirely (or at least primarily) from the high junction transparency. We emphasize that the ZBCP quantization remains unaffected at $2e^2/h$ even when the gap is considerably softened by junction tunneling effects, indicating complete preservation of Majorana topological properties.

One way of ensuring that the soft-gap phenomenon arises from the junction transparency effect (as opposed to various extrinsic processes discussed in the literature) is to go to the extreme tunneling limit with extremely high barrier height where $G_S \ll G_N$ must be satisfied, and to check that the expected functional dependence $G_S \sim G_N^2$ is obeyed in this tunneling limit. If $G_S \sim G_N^2$ is obeyed for small $G_N$, one can be confident that the soft-gap behavior for larger $G_N$ arises mainly from the junction transparency effect, thus ensuring that the Majorana properties in the topological phase would manifest non-Abelian braiding statistics. On the other hand, if the dependence of $G_S$ on $G_N$ is linear ($G_S \sim G_N$) in the extreme tunneling limit ($G_N \ll e^2/h$), then the soft-gap behavior is contaminated by physics transcending junction transparency, thus seriously hindering the topological nature of the system. The dissipation mechanism considered by us (and parametrized by the phenomenological parameter $\Gamma$ in the theory and simulation) leads to $G_S \sim G_N$ in the extreme tunneling limit, and indeed, topological properties are strongly suppressed if $\Gamma$ is the key parameter (rather than the barrier strength ($Z$), or the junction transparency ($G_N$) which is the above-gap conductance) leading to the soft-gap behavior. The same is also true when the electron temperature is large in the system, which also leads to $G_S \sim G_N$ in the strong-tunneling regime. We suggest that all tunneling experiments be automatically extended to the small $G_N$ ($\ll e^2/h$) regime to ensure the $G_S \sim G_N^2$ dependence of the induced soft gap since, otherwise, soft gap may be a serious problem for future braiding measurements.

In the context of the soft-gap behavior, experiments~\cite{Zhang2017unknown} have carried out comparison with the Beenakker formula~\cite{Beenakker1992Transport} (which is equivalent to the BTK theory~\cite{Blonder1982Transition}), and it was tentatively concluded that agreement/disagreement with the Beenakker formula connecting $G_S$ and $G_N$ is equivalent to having an effective induced hard gap in the nanowire. We have directly compared our exact numerical results with the Beenakker formula, finding the Beenakker formula, not unexpectedly, is of very limited validity. In particular, the actual regime of `safe soft gap', where the gap softness arises entirely from the BTK tunnel barrier effect (and is thus benign in spite of $G_S \ll e^2/h$ being not satisfied), is \emph{much} larger than the regime where our results agree with the Beenakker formula. In particular, our results agree with the Beenakker formula only in the high chemical potential ($\mu$) and low Zeeman splitting ($V_Z$) regime, precisely the regime where the simple analytical formula applies~\cite{Beenakker1992Transport}. We find that the induced gap, as measured by tunneling spectroscopy, may be \emph{much softer} than that implied by the Beenakker formula (i.e., $G_S$ being much larger than that implied by the Beenakker formula for a given value of $G_N$) although the gap softness arises \emph{entirely} from the junction transparency effect. In particular, in the important regime of large $V_Z$ (where the TQPT exists) the induced gap is always much softer than that given by the Beenakker formula for a given barrier strength (i.e., a given value of $G_N$) without in any way compromising the topological aspects of the system. This is a particularly important finding of our work. The induced tunneling gap being soft (or even \emph{very} soft) by itself may not be a detrimental factor in the topological nature of the nanowires as long as the softness arises from the junction transparency effect.

In contrast to the junction transparency effect causing the gap softness, temperature and dissipation causing the gap softness is a more worrisome situation. In particular, if the softness arises from dissipation, the system would be seriously compromised with respect to non-Abelian topological properties since dissipation not only suppresses the hard-gap behavior, it also suppresses the topological properties~\cite{DasSarma2016How}.  Temperature effectively suppresses the induced gap (thus, producing a soft-gap behavior) only when it is comparable to the gap itself (when $T \ll \Delta$, the thermal effect on the gap suppression is exponentially small), and near the TQPT (or for small induced gap values) this could be a serious problem. It is important to check the temperature effect experimentally by carrying out systematic temperature dependence of the soft-gap behavior, which, somewhat surprisingly, has not been much performed in the existing experiments. We emphasize that the MZMs and the associated topological properties are emergent in the Majorana nanowires only \emph{after} the system goes through the TQPT, i.e., for $V_Z>V_{Zc}$, with the induced gap vanishing at the TQPT by definition. Thus, the topological behavior of Majorana nanowires is by necessity a `\emph{small}' gap phenomenon since it happens only for large $V_Z$ where the induced gap is necessarily small.  This makes both dissipation and temperature particularly important parameters to worry about since above the TQPT, the induced gap being small, any finite temperature and dissipation become large nonperturbative parameters! Thus, our work, while establishing the fact that a soft gap by itself is not necessarily a problem for topological properties, also points to the need for producing a large induced gap above the TQPT so that any remnant dissipative and thermal effects are necessarily negligible.  To the best of our knowledge, a large induced gap in the regime where the ZBCP exists has not yet been achieved in the laboratory. Efforts should be made in this direction.

We find that the soft-gap behavior arising from junction transparency is closely tied to the broadening of MZM induced ZBCPs--in particular, when the gap is soft, the ZBCPs tend to cover most of the gap as seen experimentally almost universally. Thus, a broad ZBCP is not necessarily a problem as long as we can be sure that it arises mainly from the junction transparency effect rather than from temperature and/or dissipation effects. The height of the MZM-induced ZBCP is unaffected by junction transparency and remains at the quantized value of $2e^2/h$ (at zero temperature and with no dissipation) \emph{independent} of the background subgap conductance arising from the soft-gap behavior (for our single subband model). Thus, the ZBCP could be tiny in height above the background when the NS junction has a low tunnel barrier since then the background subgap conductance itself may be comparable to $2e^2/h$. Finite temperature and dissipation change this result considerably and nonuniversally as shown in our results (see Sec.~\ref{sec:semiinfinite} and \ref{sec:realistic}). It is important to emphasize that the MZM-induced ZBCP can never exceed $2e^2/h$ in the single-subband situation no matter how soft the gap might be. We emphasize that the often-experimentally-observed behavior of a broad ZBCP arising from essentially no discernible background gap (and covering the whole gap region) arises naturally in our results simply from the junction transparency effect although the ZBCP height is nonuniversal depending on temperature and dissipation.  The fact that the ZBCP rarely has a value of $2e^2/h$ indicates that temperature and dissipation are likely playing a role in the experimental systems in addition to the junction transparency effect.

One key difference between junction transparency and temperature/dissipation effects in producing soft gap and broadened ZBCP is that the temperature and/or dissipation preserve the area of the ZBCP as it broadens it (thus, the ZBCP height necessarily goes down as the ZBCP broadens) whereas junction transparency simply broadens the ZBCP without affecting its height (thus necessarily increasing the effective area). This difference could perhaps be used as an empirical test to find out the relative importance of BTK transparency with respect to temperature/dissipation in softening the induced gap and broadening the zero-bias peak. We note that it has been shown already~\cite{DasSarma2016How} that as long as the ZBCP height remains above $e^2/h$, the system manifests topological behavior (we have verified this explicitly too), and therefore, junction transparency by itself does not compromise topological behavior even when it leads to a very soft gap, but temperature/dissipation do compromise topological properties if they are strong enough to suppress the ZBCP height below $e^2/h$. We do emphasize, however, the serious problem of the induced gap being always very small at and above the TQPT since the gap must vanish at the TQPT.  Thus, temperature/dissipation are always important perturbations in the topological phase (above TQPT) unless the induced topological gap can be made large.

Finally, we find that the ZBCP oscillations in finite wires, arising from MZM overlaps from the two wire ends, are immune to the junction transparency effect. At zero temperature and in the absence of dissipation, ZBCP oscillations must be present in the system above the TQPT independent of how soft the gap is and how transparent the tunnel barrier might be. The presence of finite temperature and/or finite dissipation, however, suppresses the ZBCP oscillations considerably, and may even eliminate them completely if temperature and dissipation are not too small. This raises a conundrum since ZBCP oscillations are rarely observed experimentally and the soft gap is ubiquitous (at least at finite values of $V_Z$). The absence of ZBCP oscillations implies that temperature and dissipation cannot be negligible since tunnel barrier by itself cannot eliminate MZM oscillations. On the other hand, temperature and dissipation both lead to a softening of the gap independent of the junction transparency effect, and therefore, unless ZBCP oscillations are clearly seen, we cannot be sure that the soft gap physics arises only from the junction transparency physics. It therefore appears that junction transparency, temperature, and dissipation are all present in the actual experimental systems, and careful experimental work is necessary to distinguish their relative quantitative effects with respect to the phenomenology of soft gap, zero-bias peak height, and zero-bias peak oscillations. This is particularly important in view of the fact that the soft gap arising purely from the junction transparency is not a problem (for topological behaviors), but the presence of high effective temperature and dissipation (compared with the induced topological gap) is a serious problem for topological properties.


\begin{acknowledgements}
We thank Hao Zhang and Leo Kouwenhoven for helpful discussions. This work is supported by Microsoft, JQI-NSF-PFC, and LPS-MPO-CMTC. J.D.S. acknowledges the funding from Sloan Research Fellowship and NSF-DMR-1555135 (CAREER).
\end{acknowledgements}





\bibliography{BibMajorana}


\end{document}